\documentclass[a4paper,11pt]{article}
\pdfoutput=1 
\usepackage{jcappub} 

\usepackage[T1]{fontenc} 

\usepackage{graphicx}	
\usepackage{amssymb}	
\usepackage{aas_macros} 

\usepackage{orcidlink}
\usepackage{doi}

\usepackage{bm}
\usepackage{multirow}
\usepackage{physics}
\usepackage{cancel}

\usepackage{tikz}
\usetikzlibrary{shapes.geometric, arrows}

\tikzstyle{start} = [rectangle, rounded corners, minimum width=3cm, minimum height=1cm,text centered, draw=black, fill=red!30]
\tikzstyle{process} = [rectangle, minimum width=3cm, minimum height=1cm, text centered, draw=black, fill=yellow!30]
\tikzstyle{stop} = [rectangle, rounded corners, minimum width=3cm, minimum height=1cm,text centered, draw=black, fill=green!30]

\tikzstyle{arrow} = [thick,->,>=stealth]

\usepackage[nameinlink,noabbrev]{cleveref}
\crefname{equation}{Eq.}{Eqs.}
\crefname{section}{Section}{Sections}
\crefname{figure}{Fig.}{Figs.}
\crefname{table}{Table}{Tables}
\crefname{appendix}{Appendix}{Appendices}
\Crefname{figure}{Figure}{Figures}
\Crefname{equation}{Equation}{Equations}
\Crefname{section}{Section}{Sections}
\Crefname{table}{Table}{Tables}

\let\vec\mathbf

\DeclareMathOperator{\cov}{cov}

\newcommand{\SE}[1]{10^{#1}} 
\newcommand{\E}[1]{\hspace{-0.01in}\times\hspace{-0.02in}\SE{#1}}

\newcommand{\ihMpc}{h^{-1}{\rm Mpc}}
\newcommand{\ihGpc}{h^{-1}{\rm Gpc}}

\newcommand{\rascalc}{{\sc RascalC}}
\newcommand{\pycorr}{{\sc pycorr}}
\newcommand{\pyrecon}{{\sc pyrecon}}
\newcommand{\desilike}{{\sc desilike}}

\newcommand{\safett}[1]{\texorpdfstring{{\tt #1}}{#1}}

\newcommand{\abacus}{\safett{Abacus}}
\newcommand{\abacussecond}{\safett{Abacus-2}}

\newcommand{\ezmocks}{\safett{EZmocks}}

\newcommand{\snrescaling}{\alpha_{\rm SN}}
\newcommand{\snrescalingvec}{\bm \alpha_{\rm SN}}

\newcommand{\alphaiso}{{\alpha_{\rm iso}}}
\newcommand{\alphaap}{{\alpha_{\rm AP}}}

\newcommand{\supplementarylink}{\doi{10.5281/zenodo.10895161}}

\numberwithin{equation}{section}

\title{Semi-analytical covariance matrices for two-point correlation function for DESI 2024 data}

\author[1]{{M.~Rashkovetskyi}\orcidlink{0000-0001-7144-2349},}
\author[2]{{D.~Forero-Sánchez}\orcidlink{0000-0001-5957-332X},}
\author[3]{{A.~de~Mattia},}
\author[1]{{D.~J.~Eisenstein},}
\author[4]{{N.~Padmanabhan},}
\author[5]{{H.~Seo}\orcidlink{0000-0002-6588-3508},}
\author[6,7]{{A.~J.~Ross}\orcidlink{0000-0002-7522-9083},}
\author[8]{{J.~Aguilar},}
\author[9]{{S.~Ahlen}\orcidlink{0000-0001-6098-7247},}
\author[10]{{O.~Alves},}
\author[11,10]{{U.~Andrade}\orcidlink{0000-0002-4118-8236},}
\author[12]{{D.~Brooks},}
\author[3]{{E.~Burtin},}
\author[4]{{X.~Chen}\orcidlink{0000-0003-3456-0957},}
\author[8]{{T.~Claybaugh},}
\author[13]{{S.~Cole}\orcidlink{0000-0002-5954-7903},}
\author[14]{{A.~de la Macorra}\orcidlink{0000-0002-1769-1640},}
\author[15]{{Z.~Ding}\orcidlink{0000-0002-3369-3718},}
\author[12]{{P.~Doel},}
\author[16,17]{{K.~Fanning}\orcidlink{0000-0003-2371-3356},}
\author[8,18]{{S.~Ferraro}\orcidlink{0000-0003-4992-7854},}
\author[12,19]{{A.~Font-Ribera}\orcidlink{0000-0002-3033-7312},}
\author[20,21]{{J.~E.~Forero-Romero}\orcidlink{0000-0002-2890-3725},}
\author[22]{{C.~Garcia-Quintero}\orcidlink{0000-0003-1481-4294},}
\author[23,24,25]{{H.~Gil-Mar\'in}\orcidlink{0000-0003-0265-6217},}
\author[8]{{S.~Gontcho A Gontcho}\orcidlink{0000-0003-3142-233X},}
\author[26,27]{{A.~X.~Gonzalez-Morales}\orcidlink{0000-0003-4089-6924},}
\author[28]{{G.~Gutierrez},}
\author[6,29]{{K.~Honscheid},}
\author[30]{{C.~Howlett}\orcidlink{0000-0002-1081-9410},}
\author[31]{{S.~Juneau},}
\author[8]{{A.~Kremin}\orcidlink{0000-0001-6356-7424},}
\author[32]{{L.~Le~Guillou}\orcidlink{0000-0001-7178-8868},}
\author[33,19]{{M.~Manera}\orcidlink{0000-0003-4962-8934},}
\author[22]{{L.~Medina-Varela},}
\author[34]{{J.~Mena-Fern\'andez}\orcidlink{0000-0001-9497-7266},}
\author[35,19]{{R.~Miquel},}
\author[36]{{E.~Mueller},}
\author[14]{{A.~Muñoz-Gutiérrez},}
\author[37]{{A.~D.~Myers},}
\author[38]{{J.~Nie}\orcidlink{0000-0001-6590-8122},}
\author[27,39]{{G.~Niz}\orcidlink{0000-0002-1544-8946},}
\author[40,41]{{E.~Paillas}\orcidlink{0000-0002-4637-2868},}
\author[40,42,41]{{W.~J.~Percival}\orcidlink{0000-0002-0644-5727},}
\author[8,43,18]{{C.~Poppett},}
\author[14,44]{{A.~P\'{e}rez-Fern\'{a}ndez}\orcidlink{0009-0006-1331-4035},}
\author[45]{{M.~Rezaie}\orcidlink{0000-0001-5589-7116},}
\author[5]{{A.~Rosado-Marin},}
\author[46]{{G.~Rossi},}
\author[47,30]{{R.~Ruggeri}\orcidlink{0000-0002-0394-0896},}
\author[48]{{E.~Sanchez}\orcidlink{0000-0002-9646-8198},}
\author[44]{{C.~Saulder}\orcidlink{0000-0002-0408-5633},}
\author[8]{{D.~Schlegel},}
\author[49,10]{{M.~Schubnell},}
\author[31]{{D.~Sprayberry},}
\author[10]{{G.~Tarl\'{e}}\orcidlink{0000-0003-1704-0781},}
\author[31]{{B.~A.~Weaver},}
\author[2]{{J.~Yu},}
\author[50]{{C.~Zhao}\orcidlink{0000-0002-1991-7295},}
\author[38]{and {H.~Zou}\orcidlink{0000-0002-6684-3997}}

\affiliation{Affiliations are in \cref{sec:affiliations}}

\emailAdd{mrashkovetskyi@cfa.harvard.edu}

\abstract{
We present an optimized way of producing the fast semi-analytical covariance matrices for the Legendre moments of the two-point correlation function, taking into account survey geometry and mimicking the non-Gaussian effects.
We validate the approach on simulated (mock) catalogs for different galaxy types, representative of the Dark Energy Spectroscopic Instrument (DESI) Data Release 1, used in 2024 analyses.
We find only a few percent differences between the mock sample covariance matrix and our results, which can be expected given the approximate nature of the mocks,
although we do identify discrepancies between the shot-noise properties of the DESI fiber assignment algorithm and the faster approximation (emulator) used in the mocks.
Importantly, we find a close agreement ($\le 8\%$ relative differences) in the projected errorbars for distance scale parameters for the baryon acoustic oscillation measurements.
This confirms our method as an attractive alternative to simulation-based covariance matrices, especially for non-standard models or galaxy sample selections, making it particularly relevant to the broad current and future analyses of DESI data.
}


\begin{document}
\maketitle
\flushbottom



\section{Introduction}
\label{sec:intro}

It is a particularly exciting time for observational cosmology due to the transition from Stage III to Stage IV dark energy experiments.
The Dark Energy Spectroscopic Instrument (DESI) \cite{DESI2016a.Science,DESI2022.KP1.Instr} belongs to this newer generation and is actively operating.
Last year saw the validation of its scientific program \cite{DESI2023a.KP1.SV} and the early data release \cite{DESI2023b.KP1.EDR}.
As we are writing this paper in 2024, a larger 1-year dataset (DR1) \cite{DESI2024.I.DR1} is being released,
along with two-point clustering \cite{DESI2024.II.KP3},
inverse distance ladder measurements (and thus the expansion history of the Universe) using the baryon acoustic oscillations (BAO) of galaxies, quasars \cite{DESI2024.III.KP4},
and Lyman-$\alpha$ \cite{DESI2024.IV.KP6},
full-shape analysis of the 2-point statistics for galaxies and quasars constraining the growth of cosmic structure \cite{DESI2024.V.KP5},
and implications for cosmological models \cite{DESI2024.VI.KP7A,DESI2024.VII.KP7B,ChaussidonY1fnl}.
These DESI results have unprecedented precision for their kind of measurement, providing unique new opportunities to test our understanding of the Universe.

Data interpretation with a physical model requires a covariance matrix model, which can be hard to obtain.
An intuitive way to do so is from the scatter in repeated independent identical measurements.
However, only one Universe is accessible for us to probe in cosmology.
Moreover, it is not feasible to replicate a state-of-the-art experiment exactly.
Therefore covariance matrix estimation in cosmology requires elaborate techniques.

The standard method for computing covariance matrices in large-scale structure studies has been based on scatter within large sets of simulated (mock) catalogs (e.g. \cite{EZmocks,EZmocks2021,KP3s8-Zhao,Uchuu-GLAM-BOSS,FastPM-DESI}).
They need to be both highly accurate representations of the data (in particular, large enough to cover the survey volume), and numerous enough to give a good estimate of the covariance matrix of the vector of observables.
The relative precision is primarily determined by the number of samples and the dimension of the matrix.
If the number of samples is smaller than the number of bins plus one, the resulting covariance matrix estimate is not invertible.
Moreover, sample noise biases the estimate of the inverse covariance matrix \cite{hartlap-factor} and causes the widening of model parameter errorbars \cite{percival-factor-2021}.
Because highly detailed simulations require a lot of time even for volumes much smaller than a survey like DESI covers, it is unavoidable to rely on approximations, limiting the realism of the simulated catalogs.
Even so, generating and calibrating an adequate mock suite is very hard and expensive.

The mock-based covariance matrix estimation is becoming only more challenging with time.
First, as the surveys improve, each mock catalog needs to include more galaxies and/or more volume, thereby taking longer to generate and process.
Second, with more data, we aim to include a longer vector of observables in the analysis, requiring a larger covariance matrix for it, which in turn demands a higher number of mocks for adequate precision \cite{hartlap-factor,cov-matrix-accuracy,percival-factors}.
Third, the substantial time to generate mocks typically means that one cannot produce enough simulations for many separate sets of cosmological, galaxy-halo connection models and selections of tracer galaxy samples.
This creates a potential systematic error when extrapolating the covariances derived in one scenario to another.
Fourth, as a specific example of this, the long timeframe to generate mocks can even create a situation where schedule concerns force the mocks to be calibrated on early inputs that do not match the final version of the analysis.
The problem is especially severe when blinding is employed, and the simulation teams should not see the true and complete data clustering before the analysis methodology is frozen.
This creates a need and opportunity for faster alternative methods.

A promising alternative is a theoretical derivation of the covariance matrix, which is more conveniently performed in Fourier space for cosmological clustering.
Analytical covariance matrices for galaxy power spectrum have been developed using perturbation theory \cite{CovaPT,PowerSpecCovFFT,KP4s8-Alves}.
However, perturbative expansions only work for a limited range of wavenumbers.
Unfortunately, the translation of these results to the configuration space introduces an additional precision loss due to the nonlocality of the Fourier transform.
The accuracy of the transformation can be improved with the generalization of the FFTLog algorithm to covariance matrices \cite{2D-FFTLog}, but this approach has not yet been applied to and validated for redshift-space clustering of galaxies in 3 dimensions.
Therefore the analytical methods are not yet directly applicable to the correlation functions.

Another approach is re-sampling the data and considering scatter between different parts for the covariance estimation.
The jackknife technique is based on this idea.
Unfortunately, dividing the spectroscopic survey volume into equivalent parts is challenging because of boundary effects and multiple other factors varying across the survey.
This problem becomes more severe with a larger number and accordingly smaller size of regions.
More samples are still desirable for higher covariance matrix precision, analogously to the mock case.
In addition, regions are not completely independent from each other.
\cite{MP21} attempted to correct for the last factor by dividing the jackknife pair counts into different categories and re-weighting some of them.
However, \cite{fitted-jk} show that some of the former assumptions are violated in higher-density setups leading to significant biases.
To solve this problem and obtain a more precise, better-conditioned matrix, they propose a hybrid approach, combining jackknife with mocks (requiring fewer realizations than for the sample covariance).
This illustrates the promise of hybrid approaches incorporating different techniques.

Here we focus on a combination of analytical and jackknife elements in configuration space.
The theoretical component hinges on the relation between the covariance matrix of the 2-point correlation function (2PCF) and the 4-, 3- and 2-point functions.
This method has been developed in a series of papers \cite{rascal,rascal-jackknife,rascalC,rascalC-legendre-3} and implemented as the \rascalc{} code\footnote{\url{https://github.com/oliverphilcox/RascalC}}.
This approach enables the computation of covariance matrices for correlation functions using only the data, without extra effort, assumptions and approximations for generating mocks.

Multipole (Legendre) moments of the correlation function are preferable for redshift-space analysis and theoretical modeling (e.g. \cite{using-CF-multipoles-SDSS-LRG}).
\cite{rascalC} only developed an estimator for the covariance matrix of the 2PCF in angular bins.
With a large number of angular bins required to estimate multipoles adequately, the computation time of the covariance matrix becomes infeasibly long.
\cite{rascalC-legendre-3} introduced a direct covariance model for Legendre moments, but its final calibration depends on a separate computation for angular bins, which is inconvenient in practice.
We extend the methods of \cite{rascalC,rascalC-legendre-3}, developing a more practical and exact estimator for the 2PCF multipole covariance compatible with DESI's correlation function measurement code, \pycorr{}\footnote{\url{https://github.com/cosmodesi/pycorr}} \citep{pycorr}.

The method has been closely integrated with DESI.
It has been successfully applied and validated for the early BAO analysis \cite{BAO.EDR.Moon.2023,RascalC-DESI-M2}.
Following that, it was embraced as part of a coordinated covariance matrix effort for DESI DR1 two-point clustering measurements, along with analytical covariance matrices for power spectra \cite{KP4s8-Alves} and the general comparison focusing on the consistency of the model fits \cite{KP4s6-Forero-Sanchez}.
We have benefited enormously from synergies with other supporting studies for the galaxies and quasars BAO \cite{DESI2024.III.KP4}:
optimal reconstruction \cite{KP4s3-Chen,KP4s4-Paillas},
combined tracers \cite{KP4s5-Valcin},
halo occupation distribution systematics \cite{KP4s10-Mena-Fernandez,KP4s11-Garcia-Quintero},
fiducial cosmology systematics \cite{KP4s9-Perez-Fernandez}
and
theoretical systematics \cite{KP4s2-Chen}.

Fiber assignment incompleteness effects are a big challenge with DESI.
For a spectrograph with 5000 robotic fibers, the choice of a target for each is highly complex.
Some fibers cannot reach all targets in the field of view, which leads to incomplete coverage, dependent on the priority of different targets and the density of the targets on the sky.
To improve the completeness, DESI is scheduled to make 7 dark-time and 4 bright-time passes over each area during its full 5-year program \cite{SurveyOps.Schlafly.2023}.
However, the coverage achieved during the first year of observations (DR1) varies across the footprint \cite{KP3s15-Ross} between complete and only a single pass in many areas.
In multi-pass regions, the assignment also depends on previous DESI exposures in the area.
These effects imprint on 2-point clustering \cite{KP3s6-Bianchi} and must also affect the higher-point correlations entering the covariance.
The exact algorithm can be applied to mocks, but it is not fast enough to obtain as many catalogs ($\sim 1000$) as needed for the estimation or high-quality validation of the full covariance matrix \cite{KP3s7-Lasker}.
A faster approximation has been developed \cite{KP3s11-Sikandar}, enabling a suite of 1000 mocks with fiber assignment modeled.
We aim to validate the semi-analytical covariance matrices taking fiber assignment incompleteness into account for the first time.

We organize this paper in the following way.
\cref{sec:previous-work} reviews the previous results that we build upon: methodology of covariance matrix estimation in \cref{sec:cov-estimation-review}, covariance matrix comparison in \cref{sec:cov-comparison-methods}, and projection to a space of physical model parameters in \cref{sec:param-space-fisher}.
In \cref{sec:cov-estimation-new} we derive the new estimators for covariance matrices of Legendre multipoles of the correlation function, used in the remainder of the paper.
\cref{sec:simulations-methods} provides details on the mock catalogs, fiber assignment incompleteness modeling, standard BAO reconstruction and analysis techniques informing our comparison of covariance matrices.
\cref{sec:cov-setup} explains the setup for semi-analytic covariance matrix estimation.
In \cref{sec:consistency-checks} we check the consistency of the method's application to the simulations:
intrinsic numerical stability in \cref{sec:runtime-intrinsic-convergence}
and
the values of the key parameter of the covariance matrix model in \cref{sec:shot-noise-rescaling-values}.
In \cref{sec:cov-comparison-results} we compare our semi-analytical covariance matrix estimates with the sample covariance of the mocks in terms of correlation function multipoles in \cref{sec:cov-comparison-obs} and the cosmological parameters in \cref{sec:cov-comparison-param}, following both BAO and full-shape analyses.
\cref{sec:conclusion} concludes the main text with a summary and an outlook.
\cref{sec:cov-estimation-extra} gives the covariance matrix estimators generalized for multi-tracer analysis.
\cref{sec:cov-comb} explains the procedure to obtain the covariance for the combination of two disjoint regions (volumes).

\section{Overview of previous work}
\label{sec:previous-work}

\subsection{Estimation of covariance matrix for the two-point correlation functions}
\label{sec:cov-estimation-review}

\begin{figure}[tb!]
    \centering
    \includegraphics[scale=1]{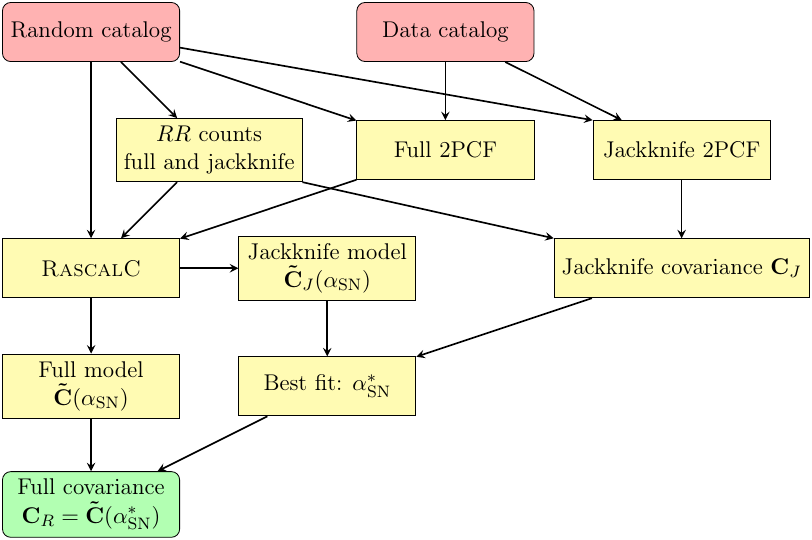}
    \caption{Flowchart of \rascalc{} jackknife pipeline (fiducial; developed in \cite{rascal-jackknife,rascalC}).
    This process is used for DESI data and most of the mock tests in this paper.
    In the latter case, a single mock catalog and its corresponding random catalog(s) are provided as data and randoms.}
    \label{fig:pipeline-jack}
\end{figure}

\begin{figure}[tb!]
    \centering
    \includegraphics[scale=1]{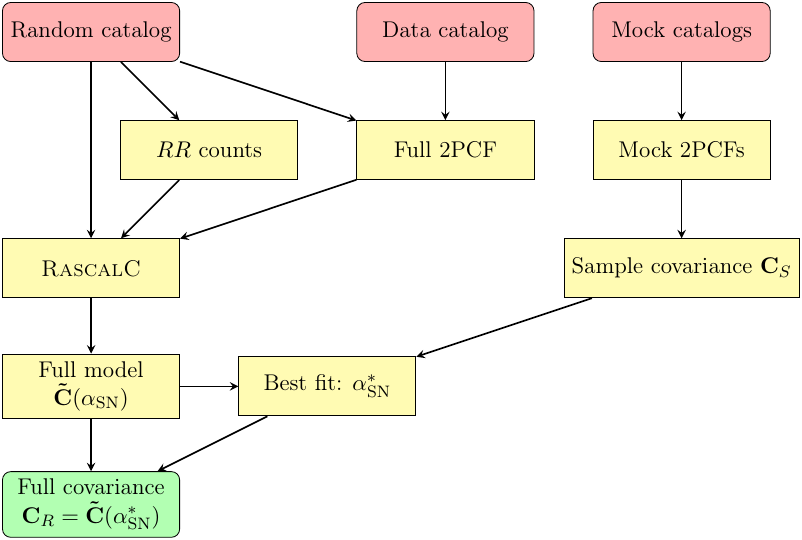}
    \caption{Flowchart of \rascalc{} mocks pipeline (alternative; principal idea from \cite{rascal}).
    Here, it is only used for additional tests on mocks (in \cref{sec:shot-noise-rescaling-values}).
    In this case, a single mock catalog and its corresponding random catalog(s) are provided as data and randoms.
    The full covariance model can be reused from a jackknife run (\cref{fig:pipeline-jack}), provided that the randoms and full 2PCF were the same.}
    \label{fig:pipeline-mocks}
\end{figure}

We start with a summary of the methodology developed in \cite{rascal,rascal-jackknife,rascalC,RascalC-DESI-M2}.
The \rascalc{} code builds single-parameter\footnote{For single tracer; for multiple tracers it is one parameter per tracer.} covariance matrix models\footnote{This step is the most computationally heavy and is implemented in C++.} based on a random catalog and a table of 2-point correlation function values.
Then we fit a model to a reference covariance (defined later) to obtain the optimal parameter value for the final prediction.
In the fiducial data pipeline, shown schematically in \cref{fig:pipeline-jack}, we measure the correlation function directly from the data, the code produces separate models for full and jackknife covariance matrices, we fit the latter to the data jackknife covariance matrix, and plug the resulting optimal parameter into the full model.
\cref{fig:pipeline-mocks} shows an alternative pipeline where we use the best fit of the full covariance model to the mock sample covariance instead.
The jackknife methodology has a significant advantage: it only requires mocks for the initial validation, and then allows to generate covariance matrices for different setups without additional simulations.
We now provide more details on both pipelines, focusing on the single-tracer case; the generalization to multiple tracers can be found in \cref{sec:cov-estimation-extra}.

To explain the covariance matrix model, we need to start with the definition of the 2-point correlation function (2PCF).
The standard Landy-Szalay estimator \cite{Landy-Szalay} in radial bin $a$ and angular bin $c$\footnote{\rascalc{} assumes uniform binning in $\abs{\mu}$, $\mu$ being the cosine of the angle between the line of sight and the pair separation (assuming symmetry with respect to $\mu \rightarrow -\mu$). The line of sight is the direction to the midpoint of the galaxy pair in an aperiodic survey, and a fixed coordinate direction ($\hat z$) in periodic boxes.} is
\begin{equation} \label{eq:Landy-Szalay-binned}
\hat\xi_a^c = \frac{\qty(NN)_a^c}{\qty(RR)_a^c}
\end{equation}
where $N=D-R$, $R$ are random points and $D$ are data points (galaxies).

The random counts $RR$ are determined by the survey geometry.
We assume that the survey design choices are independent of the random realization of the Universe.
Consequently, we treat the random counts as fixed.

The numerator, however, is determined by structure formation processes, which are considered stochastic in the cosmological paradigm.
Therefore it is crucial for scatter in the correlation function measurements.

The $NN$ counts can be expanded as
\begin{equation} \label{eq:NN-RR-discrete}
\qty(NN)_a^c = \sum_{i\neq j}n_i n_j w_i w_j \Theta^a(r_{ij}) \Theta^c(\mu_{ij}) \delta_i \delta_j.
\end{equation}
The survey has been divided into cells indexed by $i$ and $j$.
$n_i$ is the ensemble average number of galaxies in the cell $i$, $w_i$ is the weight for the random point in the cell, $\delta_i$ is the fractional galaxy overdensity in the cell, $\mu_{ij}$ is the absolute value of the cosine of the angle between the line of sight and the separation vector $\vec r_{ij} = \vec r_i - \vec r_j$, $r_{ij}$ is the length of that vector, and $\Theta$ are binning functions (unity if the argument fits into the bin and zero otherwise).

Hereafter we use the following shorthand notation for the covariance matrix:
\begin{equation} \label{eq:cov2x2_def}
C_{ab}^{cd} \equiv \cov \qty[\hat\xi_a^c, \hat\xi_b^d] \equiv \ev{\hat\xi_a^c \hat\xi_b^d} - \ev{\hat\xi_a^c} \ev{\hat\xi_b^d}.
\end{equation}

This covariance matrix involves an ensemble average of 4 overdensities at up to 4 different positions.
The reason is that each correlation function estimator (\cref{eq:Landy-Szalay-binned,eq:NN-RR-discrete}) contains 2 overdensities at 2 different positions, and they are substituted into \cref{eq:cov2x2_def}.
Some of the 4 positions can be the same, resulting in only 3 or 2 distinct positions\footnote{However, all 4 cannot be the same, because the correlation at zero separation is excluded ($i\ne j$ in \cref{eq:NN-RR-discrete}).} \cite{rascal}.

We then use the following shot-noise approximation to eliminate repeated overdensities at the same position and arrive at the $N$-point correlation functions \cite{rascal}:
\begin{equation} \label{eq:shot-noise-approximation}
\qty(\delta_i)^2 \approx \frac{\snrescaling}{n_i} \qty(1+\delta_i).
\end{equation}
The original motivation \cite{rascal} refers to the Poisson sampling.
We repeat it with slight modifications because the approximation is crucial to the method.
Assuming similar weights for galaxies and randoms, the overdensity in cell $i$ is
\begin{equation} \label{eq:cell-overdensity}
    \delta_i = \frac{b_i}{n_i} - 1,
\end{equation}
where $b_i$ is the actual number of galaxies in the cell and $n_i=\ev{b_i}$ is the expectation value (ensemble average) of that number\footnote{$n_i$ can vary from cell to cell.}.
We can choose a sufficiently small cell size so that $n_i\ll 1$.
Assuming that the galaxies appear in the cell independently, $b_i$ follows the Poisson distribution, and then $\ev{b_i^2}=n_i$.
Substituting \cref{eq:cell-overdensity} into \cref{eq:shot-noise-approximation} and taking the ensemble average of both sides gives $1/n_i - 1 \approx \snrescaling/n_i$.
This sets the baseline expectation for $\snrescaling=1$.

We further explain the overdensity $\delta_i$ in the right-hand side of \cref{eq:shot-noise-approximation}, because it cancels in the previous argument.
With $n_i\ll 1$, the average of $\qty(\delta_i)^2$ is dominated by cells with $b_i=1$.
Whereas almost all the cells have $b_i=0$, the corresponding $\qty(\delta_i)^2=1$ (\cref{eq:cell-overdensity}).
The fraction of cells with $b_i=1$ is only $\approx n_i$, but they have a large $\qty(\delta_i)^2 \approx 1/n_i^2$.
For $b_i=2$, the fraction drops significantly to $\approx n_i^2$, whereas the $\qty(\delta_i)^2 \approx 4/n_i^2$ does not increase as much.
The contributions to the average of $\qty(\delta_i)^2$ (fraction of cells times the $\qty(\delta_i)^2$ value) decrease further for higher values of $b_i$.
Thus substituting the most significant case, $b_i=1$, into \cref{eq:cell-overdensity,eq:shot-noise-approximation} gives $1/n_i^2 - 2/n_i + 1 \approx \snrescaling/n_i^2$.
This is again true for $\snrescaling=1$ because $n_i\ll 1$.
The right-hand side of \cref{eq:shot-noise-approximation} is dominated by $\delta_i\gg 1$ for $b_i=1$, therefore this overdensity is necessary.

Several factors can cause the shot-noise rescaling, i.e. effectively shift $\snrescaling$ from 1.
First, the observations of galaxies in the same small volume of a real survey are not independent.
There are fundamental limitations to their number due to resolution, number of observations, and number and size of optical fibers in a fiber spectroscopic instrument.
Furthermore, in DESI each fiber is attached to a robotic positioner confined to a specific area of the focal plane \cite{FocalPlane.Silber.2023}.
The selection of a target for the fiber (fiber assignment) therefore must depend on other objects present within this patrol area.
As a result, the number of observed galaxies $b_i$ is not guaranteed to follow the Poisson distribution and $\ev{b_i^2}=n_i$ may not hold.
Second, different weights on galaxies and random points can alter the cell overdensity estimate $\delta_i = b_i/n_i - 1$.
Such weighting is introduced, in particular, for the mitigation of fiber assignment effects on DESI clustering measurements \cite{KP3s6-Bianchi,KP3s15-Ross}.
The combination of these effects can decrease or increase the shot-noise rescaling.

To recapitulate, the shot-noise approximation (\cref{eq:shot-noise-approximation}) allows us to remove the repeated same-cell overdensities from the covariance matrix estimator.
The ensemble averages of a product of overdensities at $N$ different positions are the $N$-point correlation function by definition.
Therefore the resulting expression for the covariance matrix involves certain sums with the 4-, 3- and 2-point correlation functions.
We separate the model covariance matrix into three parts:
\begin{equation} \label{eq:Cov2x2estimator}
\tilde C_{ab}^{cd} \qty(\snrescaling) = {^4C}_{ab}^{cd} + \snrescaling {^3C}_{ab}^{cd} + \snrescaling^2 {^2C}_{ab}^{cd}.
\end{equation}
These parts, or the $d$-point terms $^d {\bf C}$ have the following theoretical expressions\footnote{In the \rascalc{} code $^d {\bf C}$ are estimated using Monte Carlo importance sampling of points from random catalogs \cite{rascalC}.}:
\begin{align} \label{eq:Cov2x2_234_Point_Defs}
{^4C}_{ab}^{cd} &= \frac{1}{\qty(RR)_a^c \qty(RR)_b^d} \sum_{i\neq j\neq k\neq l} n_i n_j n_k n_l w_i w_j w_k w_l \Theta^a(r_{ij}) \Theta^c(\mu_{ij}) \Theta^b(r_{kl}) \Theta^d(\mu_{kl}) \\ \nonumber
& \times \qty[\cancel{\eta^{({\rm c})}_{ijkl}} + 2 \xi_{ik} \xi_{jl}] \\ \nonumber
{^3C}_{ab}^{cd} &= \frac{4}{\qty(RR)_a^c \qty(RR)_b^d} \sum_{i\neq j\neq k} n_i n_j n_k w_i w_j^2 w_k \Theta^a(r_{ij}) \Theta^c(\mu_{ij}) \Theta^b(r_{jk}) \Theta^d(\mu_{jk}) \\ \nonumber
& \times \qty[\cancel{\zeta_{ijk}} + \xi_{ik}] \\ \nonumber
{^2C}_{ab}^{cd} &= \frac{2\delta^{ab}\delta^{cd}}{\qty[\qty(RR)_a^c]^2} \sum_{i\neq j} n_i n_j w_i^2 w_j^2 \Theta^a(r_{ij}) \Theta^c(\mu_{ij}) \qty[1+\xi_{ij}],
\end{align}
where $\delta^{ab}$ and $\delta^{cd}$ are Kronecker deltas; $\xi_{ij} = \xi\qty(r_{ij}, \mu_{ij})$ is the 2PCF evaluated\footnote{To compute $\xi_{ij}$, the \rascalc{} code builds a bicubic interpolator based on an input grid of $\xi\qty(r, \mu)$ values.} at the separation between points number $i$ and $j$.

$\zeta_{ijk}$ and $\eta^{{(\rm c)}}_{ijkl}$ are the 3-point and connected 4-point correlation functions.
They are evaluated at the separations between $i,j,k$ and $i,j,k,l$ points, respectively.
These non-Gaussian higher-point functions are included in the theoretical expression for completeness.
However, evaluating them in practice is challenging.
Theoretical models may not cover the necessary range of scales and might also require additional assumptions.
Direct measurements from data are costly and noisy because the number of bins increases for higher-point correlation functions.
Consequently, the 3- and connected 4-point correlation functions are not used in the practical implementation.

Instead of evaluating them, the covariance matrix has been adjusted with the shot-noise rescaling parameter according to \cref{eq:Cov2x2estimator}.
The $d$-point terms have been computed solely with the 2-point function (as designated by crossing out $\zeta$ and $\eta$ in \cref{eq:Cov2x2_234_Point_Defs}).
This 2PCF may, however, include non-linear effects, e.g. due to being estimated from the data or a realistic simulation.


The key shot-noise rescaling parameter value can be chosen for the best match to a reference covariance obtained from a set of mocks.
The best match is quantified by minimal Kullback-Leibler divergence\footnote{Which is determined by the covariance matrices for two multivariate normal distributions with the same mean.}:
\begin{equation} \label{eq:D_KL-fit-mocks}
    D_{\rm KL} \qty[{\bf \tilde C}^{-1}\qty(\snrescaling), {\bf C}_S] = \frac12 \qty[\tr({\bf \tilde C}^{-1}\qty(\snrescaling) {\bf C}_S) - N_{\rm bins} - \ln \det({\bf \tilde C}^{-1}\qty(\snrescaling) {\bf C}_S)].
\end{equation}
In its computation, the model covariance ${\bf \tilde C}\qty(\snrescaling)$ (with elements given by \cref{eq:Cov2x2estimator}) is inverted because the sample covariance of the simulations ${\bf C}_S$ is more noisy.
With a smooth theoretical model and only one parameter ($\snrescaling$) to adjust, fewer mocks are required than for direct use of their sample covariance matrix \cite{rascal}.
The process is illustrated in \cref{fig:pipeline-mocks}.
In other words, fitting the \rascalc{} covariance model to mocks is akin to template-based smoothing of the simulation-based sample covariance matrix.

Theoretically, the inversion of the \rascalc{} covariance gives a slightly biased estimate of the precision matrix.
However, the Hartlap factor \citep{hartlap-factor} is not applicable since it is not a sample covariance.
The relevant correction matrix accounting for importance sampling noise has been worked out \cite{rascal-jackknife}, but we find it practically insignificant: the eigenvalues deviate from 1 by $\lesssim 10^{-3}$.

Unfortunately, the mock-fitting approach does not solve all issues with the mocks.
The simulations still take extra time to produce and impose further assumptions and approximations.
Therefore a different method is desirable to reduce the dependence on simulations.


An alternative reference covariance can be obtained from the data with jackknife resampling.
However, the jackknife covariance matrix is not perfectly representative of the true, full-survey covariance for several reasons.
First, the resampled pieces of realistic data have different geometry from the full dataset and each other, which affects the covariance in a complicated way.
Second, the pieces of the data are correlated \cite{MP21,fitted-jk}.
Therefore it is safer to develop a separate theoretical model for jackknife covariance \cite{rascal-jackknife}, which we briefly explain in the following.

The method uses a slightly non-standard formalism, dubbed {\it unrestricted jackknife} \cite{rascalC}.
There the jackknife correlation function estimate $\xi_A$ is not the auto-correlation of the whole survey excluding the jackknife region $A$ (as in exclude-one, {\it restricted} jackknife), but the cross-correlation function between that region and the whole survey.
Equivalently, this means that the additional jackknife weighting factor for the pair of points $i,j$, $q^A_{ij}$, is 1 if both of them belong to the jackknife region $A$, $1/2$ if only one and 0 if neither.
The pair counts can be converted from different terms (auto and cross jackknife counts) often saved separately in light of Mohammad-Percival correction \cite{MP21}.

The unrestricted jackknife is convenient since the full pair counts of each type are the sum of all the jackknife pair counts of the same type.
Then if one weights the regions by the $RR$ pair counts,
\begin{equation} \label{eq:jack-weights}
\qty(w_A)_a^c = \frac{\qty(RR_A)_a^c}{\qty(RR)_a^c},
\end{equation}
the weighted mean correlation function is equal to the full-survey one.
This simplifies the theoretical modeling of the jackknife covariance.

The data jackknife covariance estimate is then
\begin{equation} \label{eq:cov-jackknife-def}
\qty(C_J)_{ab}^{cd} = \frac{\sum_A \qty(w_A)_a^c \qty(w_A)_b^d \qty[\qty(\hat\xi_A)_a^c - \hat\xi_a^c] \qty[\qty(\hat\xi_A)_b^d - \hat\xi_b^d]}{1 - \sum_A \qty(w_A)_a^c \qty(w_A)_b^d},
\end{equation}
the corresponding theoretical estimate, $\qty(\tilde C_J)_{ab}^{cd}$, is constructed analogously to \cref{eq:Cov2x2estimator}:
\begin{equation} \label{eq:Cov2x2estimator-jack}
\qty(\tilde C_J)_{ab}^{cd} \qty(\snrescaling) = \qty(^4C_J)_{ab}^{cd} + \snrescaling \qty(^3C_J)_{ab}^{cd} + \snrescaling^2 \qty(^2C_J)_{ab}^{cd}.
\end{equation}
with terms defined as \cite{rascalC}:
\begin{align} \label{eq:Cov2x2_234_Point_Defs-jack}
\qty(^4C_J)_{ab}^{cd} &= \frac{1}{\qty(RR)_a^c \qty(RR)_b^d \qty[1 - \sum_A \qty(w_A)_a^c \qty(w_A)_b^d]} \sum_{i\neq j\neq k\neq l} n_i n_j n_k n_l w_i w_j w_k w_l \\ \nonumber
& \times \Theta^a(r_{ij}) \Theta^c(\mu_{ij}) \Theta^b(r_{kl}) \Theta^d(\mu_{kl}) \qty[\cancel{\eta^{({\rm c})}_{ijkl}} + \xi_{ij} \xi_{kl} + 2\xi_{ik} \xi_{jl}] \qty(\omega_{ijkl})_{ab}^{cd} \\ \nonumber
\qty(^3C_J)_{ab}^{cd} &= \frac{4}{\qty(RR)_a^c \qty(RR)_b^d \qty[1 - \sum_A \qty(w_A)_a^c \qty(w_A)_b^d]} \sum_{i\neq j\neq k} n_i n_j n_k w_i w_j^2 w_k \\ \nonumber
& \times \Theta^a(r_{ij}) \Theta^c(\mu_{ij}) \Theta^b(r_{jk}) \Theta^d(\mu_{jk}) \qty[\cancel{\zeta_{ijk}} + \xi_{ik}] \qty(\omega_{ijjk})_{ab}^{cd} \\ \nonumber
\qty(^2C_J)_{ab}^{cd} &= \frac{2\delta^{ab}\delta^{cd}}{\qty[\qty(RR)_a^c]^2 \qty{1 - \sum_A \qty[\qty(w_A)_a^c]^2}} \sum_{i\neq j} n_i n_j w_i^2 w_j^2 \Theta^a(r_{ij}) \Theta^c(\mu_{ij}) \qty[1+\xi_{ij}] \qty(\omega_{ijij})_{ab}^{cd},
\end{align}
where $\qty(\omega_{ijkl})_{ab}^{cd}$ is an additional jackknife weight tensor:
\begin{equation}
\qty(\omega_{ijkl})_{ab}^{cd} = \sum_A \qty[q^A_{ij} - \qty(w_A)_a^c] \qty[q^A_{kl} - \qty(w_A)_b^d].
\end{equation}

To sum up, in the fiducial (jackknife) pipeline (\cref{fig:pipeline-jack}), we obtain the shot-noise rescaling by fitting the model for its jackknife covariance (\cref{eq:Cov2x2estimator-jack,eq:Cov2x2_234_Point_Defs-jack}) to the data (\cref{eq:cov-jackknife-def}).
As in the mock approach, this specifically means minimizing the Kullback-Leibler (KL) divergence between the covariance matrices:
\begin{equation} \label{D_KL-fit-jack}
    D_{\rm KL} \qty[{\bf \tilde C}_J^{-1}\qty(\snrescaling), {\bf C}_J] = \frac12 \qty[\tr({\bf \tilde C}_J^{-1}\qty(\snrescaling) {\bf C}_J) - N_{\rm bins} - \ln \det({\bf \tilde C}_J^{-1}\qty(\snrescaling) {\bf C}_J)].
\end{equation}
where the model jackknife covariance ${\bf \tilde C}_J\qty(\snrescaling)$ (\cref{eq:Cov2x2estimator-jack}) is inverted \cite{rascalC}, because the data jackknife covariance matrix ${\bf C}_J$ (\cref{eq:cov-jackknife-def}) is often not invertible.
The final covariance is obtained by plugging the resulting shot-noise rescaling values into the full covariance model (\cref{eq:Cov2x2estimator}).

Finally, we need to make adjustments for standard BAO reconstruction because it modifies the correlation function estimator.
Standard BAO reconstruction procedures shift the positions of both the data and random points.
Only shifted data ($D$) is used, whereas the randoms are kept in two variants: original ($R$) and shifted ($S$).
The correlation function is estimated via the Landy-Szalay estimator (\cref{eq:Landy-Szalay-binned}) with $N=D-S$ instead of $D-R$, but still $RR$ in the denominator.
Following \cite{RascalC-DESI-M2}, in the computations of covariance matrix terms (\cref{eq:Cov2x2_234_Point_Defs,eq:Cov2x2_234_Point_Defs-jack}) for reconstructed catalogs we use
\begin{itemize}
    \item the shifted randoms $S$ in the sums (i.e. $\bm r_i, \bm r_j, \bm r_k, \bm r_l$ used for binning functions and correlation function interpolation are the shifted random positions) because they come from expanding $NN$, which now involves $S$ and not $R$;
    \item non-shifted random counts (i.e., still $RR$) for normalization;
    \item two-point correlation functions $\xi_{ij}$ etc. with a different normalization\footnote{In the code, this is implemented by renormalizing the input grid of correlation values $\xi\qty(r, \mu)$.}: having $SS$ instead of $RR$ in the denominator, i.e.
    \begin{equation} \label{eq:Landy-Szalay-binned-post-input}
    \qty(\hat\xi_{\rm in})_a^c = \frac{\qty(NN)_a^c}{\qty(SS)_a^c}.
    \end{equation}
\end{itemize}
For data jackknife covariance (\cref{eq:cov-jackknife-def}), we still use the ordinary normalization of 2PCF (\cref{eq:Landy-Szalay-binned} with $RR$ in the denominator but $N=D-S$ instead of $D-R$).

To summarise, the key assumption is that shot-noise rescaling of purely Gaussian contributions (i.e., ignoring 3-point and connected 4-point functions) can produce a realistic covariance matrix in configuration space.
A theoretical motivation is that non-Gaussian contributions primarily affect the squeezed configurations involving small-scale correlations, below the bin width for the 2-point function, therefore not distinguishable from shot noise operating on infinitesimally small scales.
The method has been empirically shown to agree well with mock-based covariances \cite{rascal,SDSS-rascal,RascalC-DESI-M2}.

\subsection{Comparison measures for covariance matrices}
\label{sec:cov-comparison-methods}

We employ the same three compact measures of covariance matrix similarity as in \cite{RascalC-DESI-M2} for detailed comparison and validation with mocks.
We invert the final \rascalc{} covariance matrix ${\bf C}_R$ and not the mock sample covariance matrix ${\bf C}_S$ because the latter is more noisy and thus less stable to inversion.
We also find this ordering more interpretable: some additional properties do not hold with the other one, as noted below.
We start by listing the quantities and briefly explaining their meaning.

\begin{enumerate}
    \item Kullback-Leibler divergence
    \begin{equation} \label{eq:D_KL-gaussian}
    D_{\rm KL} \qty({\bf C}_R^{-1}, {\bf C}_S) = \frac12 \qty[\tr({\bf C}_R^{-1} {\bf C}_S) - N_{\rm bins} - \ln \det({\bf C}_R^{-1} {\bf C}_S)],
    \end{equation}
    which is used to optimize shot-noise rescaling, and related to the log-likelihood of the sample covariance under the assumption that the \rascalc{} covariance truly describes the distribution of mock clustering measurements\footnote{This log-likelihood relation does not hold if the KL divergence is computed between sample precision (inverse covariance) and theoretical covariance matrices.} \cite{rascal}.
    \item The directional RMS relative difference:
    \begin{equation} \label{eq:R_inv-definition}
    R_{\rm inv} \qty({\bf C}_R^{-1}, {\bf C}_S) = \frac1{\sqrt{N_{\rm bins}}} \norm{{\bf C}_S^{1/2} {\bf C}_R^{-1} {\bf C}_S^{1/2} - {\bf \mathbb I}}_F = \sqrt{\frac{\tr[\qty({\bf C}_R^{-1} {\bf C}_S - {\bf \mathbb I})^2]}{N_{\rm bins}}}.
    \end{equation}
    $D_{\rm KL}$ and $R_{\rm inv}$ are sensitive to deviations in different ways, but their expectation values would not be zero even if \rascalc{} covariance matrices perfectly matched the true underlying covariance describing the distribution of mock realizations.
    \item The mean reduced chi-squared:
    \begin{equation} \label{eq:chi2_red-definition}
    \chi^2_{\rm red} \qty({\bf C}_R^{-1}, {\bf C}_S) = \frac1{N_{\rm bins}} \tr({\bf C}_R^{-1} {\bf C}_S),
    \end{equation}
    which is distributed like a reduced $\chi^2$ with $N_{\rm bins} \times (n_S-1)$ degrees of freedom under the assumption that \rascalc{} covariance describes the distribution of the mock clustering measurements\footnote{Again, if we computed the mean reduced $\chi^2$ between the mock precision and theoretical covariance matrices, it would not follow the reduced $\chi^2$ distribution exactly.}.
    This measure is more sensitive to the overall scaling between the covariance matrices, whereas deviations in different directions can cancel each other.
\end{enumerate}

It is also important to set expectations for these three comparison measures in the perfect case.
By this, we mean comparing the true precision matrix with the sample covariance estimated from $n_S$ multivariate normal samples following the same covariance.
The distribution of clustering statistics can be assumed normal.
This allows us to test the hypothesis that \rascalc{} can predict the true covariance of the mocks.

We compute the statistical means and standard deviations for our comparison measures in two ways:
\begin{itemize}
    \item Theoretically --- using the expressions derived in \cite{RascalC-DESI-M2}.
    It is important to note that the results for $\chi^2_{\rm red}$ are exact while the others are obtained with approximations.
    \item Empirically (Monte-Carlo) --- using a large number of multivariate normal samples.
    We generate $N_r=10,000$ chunks of $n_S$ independent samples of $N_{\rm bins}$-dimensional multivariate normal vectors with a (true) unit covariance matrix\footnote{The value of the true covariance matrix does not matter because it cancels out in each comparison measure (ignoring the numerical instabilities). The dimension, however, is crucially important. Therefore we repeat the procedure for each value of $N_{\rm bins}$ relevant for comparisons in this work.}.
    In each chunk, we estimate the sample covariance and compute the comparison measures between their true (unit) precision matrix and the sample covariance estimate.
    Finally, we estimate the mean and standard deviation for each comparison measure using the obtained $N_r$ random realizations\footnote{These realizations of covariance matrix comparison measures (for $n_S=1000$), as well as all estimates of their means and standard deviations (theoretical, empirical and our fiducial choice between the two for each case), are provided in the supplementary material: \supplementarylink{}.}.
\end{itemize}
We choose a preferred (fiducial) value between the theoretical estimate and the corresponding empirical figure in each case.
We prefer the theoretical value unless the difference is more than $3\sigma$.
Otherwise, we select the empirical estimate.
As suspected, we only find $>3\sigma$ deviations for KL divergence (with larger numbers of bins) and $R_{\rm inv}$ (for smaller numbers of bins).
The fiducial values are used in \cref{tab:cov-comparison-full,tab:cov-comparison-shape-range,tab:cov-comparison-BAO-range,tab:cov-comparison-BAO-parameters,tab:cov-comparison-ShapeFit-parameters,tab:cov-comparison-direct-fit-parameters} as reference for comparison measures between \rascalc{} precision matrices and sample covariances.

\subsection{Fisher projection to the space of model parameters}
\label{sec:param-space-fisher}

Comparing the full covariance matrices for observables (e.g., binned 2PCF or 2PCF multipoles) may be overly generic.
In such a case, we consider every arbitrary ``direction'' in the high-dimensional linear space of these observables.
Some of these directions may be unphysical, whereas others can have little or no connection to the data analysis.
To highlight a smaller number of meaningful, impactful directions, \cite{subspace-projection,cov-comparison-projection} suggested projecting the covariance matrices into a model-dependent subspace.

We project the covariance matrices to the model parameters through the inverse of the Fisher matrix, following \cite{RascalC-DESI-M2} for simplicity.
For \rascalc{} results, one can use a simple expression neglecting the inversion bias (Hartlap-like) corrections\footnote{As we mentioned in \cref{sec:cov-estimation-review}, such corrections are small for \rascalc{}.}:
\begin{equation}
{\bf C}_R^{\rm par} = \qty[{\bf M} \qty({\bf C^\prime}^{\rm obs}_R)^{-1} {\bf M}^T]^{-1}. \label{eq:parameter-covariance-rascalc}
\end{equation}
${\bf C^\prime}^{\rm obs}_R$ is the \rascalc{} covariance matrix restricted to the separation and multipole range of the respective model.
This range consitutes $N^\prime_{\rm bins}$ observables.
${\bf M}$ is the Jacobian of the model, i.e. the matrix of derivatives of radially-binned 2PCF multipoles $\bm \xi$ with respect to the parameter vector $\bm \theta$:
\begin{equation}
M_{pa} \equiv \frac{\partial \xi_a}{\partial \theta_p}. \label{eq:model-jacobian}
\end{equation}

For the mock sample covariance, we use a similar expression:
\begin{equation}
{\bf C}_S^{\rm par} = \frac{n_S-1}{n_S-N^\prime_{\rm bins}+N_{\rm pars}-1} \qty[{\bf M} \qty({\bf C^\prime}^{\rm obs}_S)^{-1} {\bf M}^T]^{-1}.  \label{eq:parameter-covariance-sample}
\end{equation}
$n_S$ is the number of mock samples used to estimate the sample covariance, and $N_{\rm pars}$ is the number of model parameters.

The important difference from \cref{eq:parameter-covariance-rascalc} is that the sample covariance estimate is more noisy than the semi-analytical model result.
Therefore non-linear matrix inversion operations cause significant biases.
\cite{density-split-clustering} considered this problem and gave a single (Hartlap-like) correction factor in their Eq.~(B6), which we use here.
The additional multiplier removes the inversion biases to leading order.

Unlike in \cite{RascalC-DESI-M2}, we exclude the nuisance parameters.
We leave out the corresponding rows and columns from the covariance matrices given by \cref{eq:parameter-covariance-sample,eq:parameter-covariance-rascalc}.
The remaining sub-matrix represents the covariance of the parameters of interest marginalized over the nuisance parameters.

\section{Covariance for projected Legendre moments of 2PCF revisited}
\label{sec:cov-estimation-new}

Theoretical models of large-scale structure often use multipole (Legendre) moments of the 2-point correlation function instead of its angularly ($\mu$) binned estimates (e.g. \cite{using-CF-multipoles-SDSS-LRG,KP4s2-Chen}).
Moreover, the number of multipoles of interest is typically low --- monopole, quadrupole, and sometimes hexadecapole \cite{KP5s1-Maus,KP5s2-Maus,KP5s3-Noriega,KP5s4-Lai,KP5s5-Ramirez}\footnote{These references discuss the power spectrum modeling, but the $\ell$'th Legendre moment of the correlation function $\xi_\ell$ is determined only by the same-order multipole of the power spectrum $P_\ell$ via a spherical Bessel $j_\ell$ transform.}.
It is then convenient to compress the correlation by converting the angularly-binned correlation function (with more than 3 angular bins) to the Legendre moments.
Covariance matrices for Legendre multipoles have a lower dimension, which causes fewer numerical problems and makes them easier to estimate directly.

The covariance matrix model for Legendre moments was developed in \rascalc{} previously \cite{rascalC-legendre-3}, but this implementation has several disadvantages.

First, it is not directly compatible with jackknives.
In practice, producing the Legendre moment covariance with optimal shot-noise rescaling based on data (not relying on a mock sample) requires two separate computations: one for angular ($\mu$) bins with jackknives to tune the shot-noise rescaling, and another to construct the full covariance matrix model for Legendre multipoles.
This is an inconvenience when one does not intend to use the angularly-binned correlation function in cosmological inference.

Second, the 2PCF estimation library used in DESI, \pycorr{}\footnote{\url{https://github.com/cosmodesi/pycorr}} \citep{pycorr}, operates under slightly different assumptions.
\pycorr{} uses the angularly binned 2-point correlation function $\hat\xi_a^c$ (\cref{eq:Landy-Szalay-binned}) estimates with a large but finite number ($\sim 100$) of angular bins to compute the radially binned Legendre moments:
\begin{equation}
    \hat\xi^\ell_a = \qty(2\ell+1) \sum_c \hat\xi_a^c \int_{\Delta\mu_c} d\mu\, L_\ell(\mu) = \sum_c \hat\xi_a^c F^\ell_c; \label{eq:Legendre-from-binned-2PCF}
\end{equation}
where
\begin{equation}
    F^\ell_c \equiv \qty(2\ell+1) \int_{\Delta\mu_c} d\mu\, L_\ell\qty(\mu) \label{eq:Legendre-projection-factors}
\end{equation}
are the projection factors, which do not depend on radial bins or the tracers involved in the correlation function.
The equations above assume even multipole index $\ell$, and binning in $\abs{\mu}\in [0,1]$\footnote{\pycorr{} is capable of binning in $\mu\in [-1,1]$, retaining the sign information. But all auto-correlation functions are necessary symmetric, and it is easy to ``wrap'' the counts to $\abs{\mu}$ bins in any case as long as the number of $\mu$ bins is even.}.
In contrast, the \rascalc{} estimators developed earlier assume weighting by Legendre polynomials during pair counting \cite{rascalC-legendre-3}.
This is equivalent to using infinitesimally narrow angular bins in \cref{eq:Legendre-from-binned-2PCF}.
The mismatch in assumptions between \pycorr{} and \rascalc{} may not cause significant differences in practice, but it is not desirable.

Third, an additional step was needed to account for realistic survey geometry.
The previous \rascalc{} realization relies on the survey correction function --- the ratio of pair counts in a real survey and a periodic box with the same volume \cite{survey-correction-factor-ref}.
This function needed to be modeled for arbitrary angles ($\abs{\mu}$).
A piecewise-polynomial form has been assumed \cite{rascalC-legendre-3}.
Whereas they explain the need for two different polynomials by the particularly strong redshift-space distortions near the line of sight ($\abs{\mu}=1$), the choice of the partition point\footnote{I.e. the boundary between the two polynomials, where they join continuously and smoothly.} at $\abs{\mu}=0.75$ has not been motivated and may not be the best.
This might be a minor issue as well, but still a source of extra uncertainty.

We have seen an opportunity to address all three issues and streamline the covariance matrix computation procedure for extensive usage with DESI.
Since the projection in \cref{eq:Legendre-from-binned-2PCF} is linear, the covariance matrix for these Legendre moments estimators can be obtained from the $r,\mu$-binned one given by \cref{eq:Cov2x2estimator}:
\begin{equation} \label{eq:full-cov-projection-Legendre}
    \tilde C_{ab}^{\ell\ell'} \equiv \cov \qty[\hat \xi^\ell_a, \hat \xi^{\ell'}_b] = \sum_{c,d} \tilde C_{ab}^{cd} F^\ell_c F^{\ell'}_d.
\end{equation}

A major technical result of this paper is a methodology to compute this covariance matrix of the Legendre multipoles directly at the level of the summation over point configurations, rather than having to compute and then project the much larger covariance matrix of fine angular bins.
For this, several quantities need to be inserted into the sums of \cref{eq:Cov2x2_234_Point_Defs}, and we obtain the following 4, 3, and 2-point terms:
\begin{align} \label{eq:Cov2x2_234_Point_Defs_Legendre_Projected}
{^4C}_{ab}^{\ell\ell'} &= \sum_{i\neq j\neq k\neq l} n_i n_j n_k n_l w_i w_j w_k w_l \Theta^a(r_{ij}) \Theta^b(r_{kl}) \qty[\cancel{\eta^{({\rm c})}_{ijkl}} + 2\xi_{ik} \xi_{jl}] \\ \nonumber
& \times \sum_c \frac{\Theta^c(\mu_{ij}) F^\ell_c}{\qty(RR)_a^c} \sum_d \frac{\Theta^d(\mu_{kl}) F^{\ell'}_d}{\qty(RR)_b^d}, \\ \nonumber
{^3C}_{ab}^{\ell\ell'} &= 4 \sum_{i\neq j\neq k} n_i n_j n_k w_i w_j^2 w_k \Theta^a(r_{ij}) \Theta^b(r_{jk}) \qty[\cancel{\zeta_{ijk}} + \xi_{ik}] \sum_c \frac{\Theta^c(\mu_{ij}) F^\ell_c}{\qty(RR)_a^c} \sum_d \frac{\Theta^d(\mu_{jk}) F^{\ell'}_d}{\qty(RR)_b^d}, \\ \nonumber
{^2C}_{ab}^{\ell\ell'} &= 2\delta^{ab} \sum_{i\neq j} n_i n_j w_i^2 w_j^2 \Theta^a(r_{ij}) \qty[1+\xi^{XY}_{ij}] \sum_c \frac{\Theta^c(\mu_{ij}) F^\ell_c F^{\ell'}_c}{\qty[\qty(RR)_a^c]^2}.
\end{align}
As in \cref{eq:Cov2x2_234_Point_Defs}, we include the non-Gaussian higher-point functions in these theoretical equations.
However, we drop them in the current implementation, which we signify by crossing them out.
Sums like $\sum_c \Theta^c(\mu) \dots$ practically mean finding the angular bin $\tilde c$ to which the $\mu$ value belongs and then evaluating the rest only for that one bin.
Within the code, we sample a quadruplet, triplet, or pair of particles and then accumulate its contribution to all the Legendre multipole moments in its radial bin.

These 4, 3, and 2-point terms can be combined to the full theoretical estimate analogously to \cref{eq:Cov2x2estimator}, i.e.
\begin{equation} \label{eq:cov2x2_Legendre}
\tilde C_{ab}^{\ell\ell'} \qty(\snrescaling) = {^4C}_{ab}^{\ell\ell'} + \snrescaling {^3C}_{ab}^{\ell\ell'} + \snrescaling^2 {^2C}_{ab}^{\ell\ell'}.
\end{equation}
This is the single-tracer expression; the version for multiple tracers is provided in \cref{sec:cov-estimation-multi-legendre-projected}.

For simplicity, we have decided to reuse the $r,\mu$ binned jackknife covariance matrix estimate (Eq.~\eqref{eq:cov-jackknife-def}).
An alternative could be a tedious re-derivation of the theoretical jackknife covariance model with some weights for individual jackknife multipole estimators.
The method only requires this step to calibrate the shot-noise rescaling parameter.
The precise choice of the reference jackknife covariance should not matter as long as the data and model estimators are treated consistently.
Consequently, we project the angularly-binned data jackknife covariance matrix (Eq.~\eqref{eq:cov-jackknife-def}) similarly to the full covariance (Eq.~\eqref{eq:full-cov-projection-Legendre}):
\begin{equation} \label{eq:jack-cov-projection-Legendre}
    \qty(C_J)_{ab}^{\ell\ell'} = \sum_{c,d} \qty(C_J)_{ab}^{cd} F^\ell_c F^{\ell'}_d
\end{equation}
and do the same with the theoretical prediction (\cref{eq:Cov2x2estimator-jack}):
\begin{equation} \label{eq:Cov2x2estimator-jack-legendre}
\qty(\tilde C_J)_{ab}^{\ell\ell'} \qty(\snrescaling) = \qty(^4C_J)_{ab}^{\ell\ell'} + \snrescaling \qty(^3C_J)_{ab}^{\ell\ell'} + \snrescaling^2 \qty(^2C_J)_{ab}^{\ell\ell'}.
\end{equation}
The jackknife $d$-point terms (\cref{eq:Cov2x2_234_Point_Defs-jack}) accordingly are transformed to
\begin{align} \label{eq:Cov2x2_234_Point_Defs-jack-legendre}
\qty(^4C_J)_{ab}^{\ell\ell'} &= \sum_{i\neq j\neq k\neq l} n_i n_j n_k n_l w_i w_j w_k w_l \Theta^a(r_{ij}) \Theta^b(r_{kl}) \qty[\cancel{\eta^{({\rm c})}_{ijkl}} + \xi_{ij} \xi_{kl} + 2\xi_{ik} \xi_{jl}] \\ \nonumber
& \times \sum_{c,d} \frac{\qty(\omega_{ijkl})_{ab}^{cd} \Theta^c(\mu_{ij}) \Theta^d(\mu_{kl}) F^\ell_c F^{\ell'}_d}{\qty(RR)_a^c \qty(RR)_b^d \qty[1 - \sum_A \qty(w_A)_a^c \qty(w_A)_b^d]}, \\ \nonumber
\qty(^3C_J)_{ab}^{\ell\ell'} &= 4 \sum_{i\neq j\neq k} n_i n_j n_k w_i w_j^2 w_k \Theta^a(r_{ij}) \Theta^b(r_{jk}) \qty[\cancel{\zeta_{ijk}} + \xi_{ik}] \\ \nonumber
& \times \sum_{c,d} \frac{\qty(\omega_{ijjk})_{ab}^{cd} \Theta^c(\mu_{ij}) \Theta^d(\mu_{jk}) F^\ell_c F^{\ell'}_d}{\qty(RR)_a^c \qty(RR)_b^d \qty[1 - \sum_A \qty(w_A)_a^c \qty(w_A)_b^d]}, \\ \nonumber
\qty(^2C_J)_{ab}^{\ell\ell'} &= 2\delta^{ab} \sum_{i\neq j} n_i n_j \qty(w_i w_j)^2 \Theta^a(r_{ij}) \qty[1+\xi_{ij}] \sum_c \frac{\qty(\omega_{ijij})_{ab}^{cc} \Theta^c(\mu_{ij}) F^\ell_c F^{\ell'}_c}{\qty[\qty(RR)_a^c]^2 \qty{1 - \sum_A \qty[\qty(w_A)_a^c]^2}}.
\end{align}
Similarly to \cref{eq:Cov2x2_234_Point_Defs_Legendre_Projected}, sums of the form $\sum_{c,d} \Theta^c(\mu_1) \Theta^d(\mu_2) \dots$ practically mean finding the angular bins $\tilde c,\tilde d$ to which the $\mu_{1,2}$ values belong correspondingly and then evaluating the rest ($\dots$) only for that pair of bins.
A given set of cells contributes to covariance of one pair of angular bins ($c,d$ in \cref{eq:Cov2x2_234_Point_Defs}), but all multipoles ($\ell,\ell'$).

We have also omitted the small disconnected part of the 4-point jackknife term ($\xi_{ij} \xi_{kl}$) in the main code implementation for practical reasons.
This part is estimated in $r,\mu$ bins (\cref{eq:Cov2x2_234_Point_Defs-jack}) by separating it into a product of two sums that need to be computed for each jackknife region \cite{rascalC}.
The evaluation becomes less convenient in Legendre multipoles due to the projection factors inserted into the sum over cells/particles.
We have computed the disconnected term in a couple of realistic setups by using the technique for $r,\mu$ bins\footnote{Storing more data due to keeping many (100) angular ($\mu$) bins for each of 60 jackknife regions. The number of angular bins could be reduced, but at the price of potential loss of precision.} and projecting the resulting covariance matrix part into the multipole moments (analogously to \cref{eq:jack-cov-projection-Legendre}).
We have found that the inclusion of the disconnected term does not change the shot-noise rescaling values up to the 6th digit after the decimal point.
The optimal shot-noise rescaling parameter is the only link between the disconnected jackknife term and the final covariance matrix.
Therefore we concluded that the impact of the disconnected term is practically negligible.

In addition, we give a theoretical justification for neglecting the disconnected term, although not completely strict.
The disconnected jackknife term vanishes exactly if either the correlation function is constant, the jackknife regions are identical, or the jackknife counts in each region are the same in each fine bin \cite{rascalC}.
An arbitrary survey would not meet any of these conditions exactly, but they likely hold approximately.
Therefore the disconnected term is expected to be small.

This concludes the description of the new covariance estimators we apply to DESI data and mocks.
To reiterate, the key practical advantage is that a single computation provides the Legendre covariance with shot-noise rescaling tuned on jackknives.
Additionally, the new estimators use the random-random counts directly, instead of relying on a survey correction function fit for realistic survey geometry \cite{rascalC-legendre-3}.
Therefore we use this method for covariance matrix estimation in the rest of this paper.

\section{DESI simulations and methods}
\label{sec:simulations-methods}

In this section, we briefly describe the simulations we use and their processing steps.

\subsection{Mocks}
\label{sec:mock-types}

In this work, we mainly rely on effective Zel'dovich mocks (\ezmocks{}) \cite{EZmocks,EZmocks2021}.
We use the suite of 1000 catalogs representative of DESI DR1 \cite{KP3s8-Zhao}.
These mocks are more approximate than those relying on full $N$-body simulations.
However, they are fast enough to make a large number of simulations in $6~\ihGpc$ boxes covering the whole volume of DESI DR1 without replications.

For shot-noise rescaling investigation (\cref{sec:shot-noise-rescaling-values}) we also used more realistic \abacussecond{} mocks.
They are based on the {\sc AbacusSummit} suite of $N$-body simulations \citep{AbacusSummit} produced with the {\sc Abacus} code \citep{Abacus-code}.
The halos have been identified with the {\sc CompaSO} halo finder \citep{CompaSO-halo-finder}.
The galaxy catalogs have been produced within halo occupancy distribution (HOD) formalism using the {\sc AbacusHOD} framework \citep{AbacusHOD}.
More details on the suite representing DESI DR1 can be found in \cite{DESI2024.III.KP4}.
These simulations have two disadvantages: only 25 realizations and smaller ($2~\ihGpc$) boxes requiring replications to cover the DESI DR1 volume.

The factors of number and volume are crucial for covariance matrices.
The number of mocks sets the relative precision of the sample covariance estimate, which is the only reference we have for comparison.
By replications, we mean different parts of the final galaxy catalog made from the same part of the original simulation.
These parts can become too strongly correlated and therefore bias the sample covariance estimate.
This motivates our choice of \ezmocks{} for the majority of this work.

\subsection{Fiber assignment modeling}
\label{sec:fiber-assignment-models}

An important novel aspect of this work is the application of semi-analytical covariance matrices to mocks that incorporate a model of fiber assignment effects.

We mainly use the approximate algorithm called ``fast fiber assign'' or ``fast fiber assignment'' (FFA) \cite{KP3s11-Sikandar,KP3s6-Bianchi}.
It emulates the DESI fiber assignment algorithm using less computational resources \cite{DESI2024.III.KP4}.

For shot-noise rescaling investigation (\cref{sec:shot-noise-rescaling-values}) we also used the mock realizations of the DESI fiber assignment algorithm.
They involve alternate merged target ledgers, thus the shortcut AMTL \cite{KP3s7-Lasker}.
However, this method is prohibitively slow to process all 1000 \ezmocks{} \cite{DESI2024.III.KP4}.

These fiber assignment algorithms vary in how they assign weights to galaxies and random points to mitigate incompleteness effects.
As discussed in section 5.1 of \cite{KP3s15-Ross}, the DESI assignment completeness can be split into two components, one of which can be modeled by assigning weights to randoms.
However, the FFA algorithm only determines the total assignment completeness per galaxy.
Consequently, the weights per galaxy resulting from FFA have greater variance than for AMTL mocks and the DESI DR1 LSS catalogs.
As we mentioned earlier, different weighting methods for data and random points can affect the shot-noise rescaling values (\cref{eq:shot-noise-approximation}).

Effects of fiber assignment may pose additional challenges to the method because it involves anisotropic pair-wise sampling, depending both on the density of the targets and the number of survey passes in the region \cite{KP3s6-Bianchi}.
We provide the \rascalc{} code with the random catalog and clustering estimate affected by fiber assignment (see the flowchart in \cref{fig:pipeline-jack}), but the expansion leading to \cref{eq:Cov2x2_234_Point_Defs} (and \cref{eq:Cov2x2_234_Point_Defs-jack}) uses the survey-wide correlation function(s) to calculate the ensemble averages of products of overdensities, and the shot-noise rescaling (\cref{eq:shot-noise-approximation}) is also global.
It is challenging to let the correlation function or shot-noise rescaling vary across the survey (with on-sky position or redshift) without complicating the covariance matrix model too much and introducing too many parameters.

Fiber assignment incompleteness might also cause issues with the jackknife.
Ideally, we would like each sub-region to have a distribution of the number of passes representative of the full survey.
This is challenging to achieve, and the current jackknife assignment does not guarantee that.

\subsection{Reconstruction}
\label{sec:recon}

We apply the semi-analytical covariance matrix estimators before and after standard BAO reconstruction (pre- and post-recon).
The reconstruction methodology follows the findings of the DESI DR1 optimal reconstruction task force \cite{KP4s3-Chen,KP4s4-Paillas}:
the {\tt RecSym} mode of the {\tt IterativeFFTReconstruction} 
algorithm \cite{recon-fourier-space} from the \pyrecon{} package\footnote{\url{https://github.com/cosmodesi/pyrecon}} with smoothing scale of $15\,\ihMpc$.

\subsection{Theoretical modeling and fitting}
\label{sec:theoretical-models-fits}

Our first analysis of interest is DESI 2024 baryon acoustic oscillations \cite{DESI2024.III.KP4}.
We use the same anisotropic (2D) model with BAO power spectrum template and spline-based broadband terms \cite{KP4s2-Chen}.
The fit uses monopole and quadrupole in radial bins spanning $s=48-152\,\ihMpc$.

The second analysis of interest is DESI 2024 full-shape \cite{DESI2024.V.KP5}.
This presents more difficulties because it primarily uses power spectra and the methodology has not been standardized for correlation functions.
Nevertheless, we use similar models relying on the {\tt velocileptors} Lagrangian perturbation theory model \cite{velocileptors-Chen2020,velocileptors-Chen2021,KP5s2-Maus} with maximum freedom and standard prior basis.
Two variants correspond to the power spectrum template choice.
One approach is ``ShapeFit''.
It is a compression method using parametric variations of a single power spectrum template evaluated at the reference cosmology \cite{ShapeFit-Brieden21}.
The other is ``full modeling'', meaning a linear power spectrum from \texttt{CLASS} Boltzmann code \citep{CLASS}.
The fit uses monopole, quadrupole and hexadecapole in radial bins spanning $s=28-152\,\ihMpc$.

These analyses inform our work in two ways.

First, the two fit ranges set our choice of sub-matrices of correlation function multipoles covariances for comparison.
In other words, we restrict the covariance matrices to the multipoles and radial bins used in the fitting.
This gives us an informed choice for smaller sub-matrices to see which parts of the covariance matrix are captured better by the semi-analytic method.

Second, we project the covariance matrices into parameter spaces of these three models.
We do this with the inverse Fisher matrix as described in \cref{sec:param-space-fisher}.
It is very important to see that the errorbars on physical parameters are predicted reliably through the \rascalc{} covariances, although this analysis may not generalize to alternative models.

We use the \desilike{} package\footnote{\url{https://github.com/cosmodesi/desilike}} for the evaluation of all three models (including polynomial emulators for the full-shape models) and fitting the data.

\section{Setup for semi-analytical covariance matrices}
\label{sec:cov-setup}

\begin{table}[tb]
    \centering
    \begin{tabular}{|c|c|c|c|}
        \hline
        Tracer & LRG & ELG & BGS \\
        \hline
        $z$ range & $(0.8, 1.1)$ & $(1.1, 1.6)$ & $(0.1, 0.4)$ \\
        \hline
        Designation in \cite{DESI2024.II.KP3,DESI2024.III.KP4} & {\tt LRG3} & {\tt ELG2} & {\tt BGS} \\
        \hline
        \ezmocks{} snapshot $z$ & 1.1 & 1.325 & 0.2 \\
        \hline
    \end{tabular}
    \caption{Tracers and redshift bins used in this paper.
    For LRG and ELG, which have multiple bins unlike BGS, we have selected the densest ones, as shot-noise seems easier to capture with \rascalc{}.
    We did not include the quasars (QSO) for the same reason.
    The snapshot redshifts were used to construct the power spectrum templates.}
    \label{tab:tracers-bins}
\end{table}

We consider three galaxy types (tracers) in three different redshift ranges.
We use luminous red galaxies (LRG) with $z=0.8-1.1$, emission line galaxies (ELG) \cite{ELG.TS.Raichoor.2023} with $z=1.1-1.6$ and magnitude-limited bright galaxy survey (BGS) \cite{BGS.TS.Hahn.2023} with $z=0.1-0.4$.
These correspond to {\tt LRG3}, {\tt ELG2} and {\tt BGS} samples in the main DESI BAO papers \cite{DESI2024.III.KP4,DESI2024.VI.KP7A} respectively.
The summary of tracers and redshift bins is provided in \cref{tab:tracers-bins}.

We apply the data pipeline\footnote{The data scripts are available at \url{https://github.com/cosmodesi/RascalC-scripts/tree/DESI2024/DESI/Y1}, {\tt pre} and {\tt post} directories for single-tracer covariances before and after reconstruction respectively.
Our analogous single-mock scripts (with only minor differences) are in \url{https://github.com/cosmodesi/RascalC-scripts/tree/DESI2024/DESI/Y1/EZmocks/single}, {\tt pre} and {\tt post} folders.} (\cref{fig:pipeline-jack}) to single mock catalogs, as was done in previous works.
This implies using the random files, full and jackknife correlation function estimates specific to that catalog.
We use 10\footnote{The number is a compromise between getting good statistics and saving computing time.} different realizations of \ezmocks{} to quantify the fluctuations in the semi-analytical results due to realistic random variations in the input quantities.

We process the North and South Galactic Caps (NGC and SGC) separately.
DESI DR1 data has been processed in the same manner.
This allows different shot-noise rescaling values, reflecting different completeness patterns in these parts of DESI DR1.
Then we combine the two covariances into a single matrix for the full survey assuming the Galactic Caps are uncorrelated\footnote{In the \ezmocks{} NGC and SGC are constructed from separate realizations \cite{KP3s8-Zhao}, so they are independent. As a result, we do not validate this assumption for data.} (\cref{sec:cov-comb}).

We create the covariance matrices for monopole, quadrupole and hexadecapole in 45 radial bins between 20 and 200~$\ihMpc$ (each 4~$\ihMpc$ wide).
We exclude the $s<20\,\ihMpc$ bins because they impede the convergence of the covariance matrices.
Moreover, we expect the shot-noise rescaling to become inadequate on small scales.

\section{Intrinsic tests of the method}
\label{sec:consistency-checks}

Having described the setup for the current \rascalc{} application to \ezmocks{}, we detail the quality checks we perform before comparing the results with the mock sample covariance.

\subsection{Intrinsic convergence and computation time}
\label{sec:runtime-intrinsic-convergence}

We have two specific convergence criteria for each \rascalc{} computation.
The first is the eigenvalue test performed by the code on both full and jackknife covariance matrix terms (\cref{eq:Cov2x2_234_Point_Defs_Legendre_Projected,eq:Cov2x2_234_Point_Defs-jack-legendre}).
The desired condition is that the minimal eigenvalue of the 4-point term is larger than minus the minimal eigenvalue of the 2-point term \cite{rascalC}.
The second is the positive definiteness of the final covariance matrix estimate.
We do not use the covariance matrix model that fails either of these tests.

In addition, we have a quantitative measure of convergence without strict thresholds.
As in \cite{RascalC-DESI-M2}, we use the $R_{\rm inv}$ comparison measure (\cref{eq:R_inv-definition}) between the different estimates of each \rascalc{} covariance matrix from separate halves of the Monte-Carlo integration samples.
We found several high outliers compared to other mock realizations of the same tracer.

We repeat the computation on the mocks with the convergence issues mentioned before.
Then we post-process the data from the second computation and the data combined from the two computations.
We choose the one that first satisfies the two strict criteria and then gives a lower $R_{\rm inv}$ value.

After this, we find that LRG covariance matrices reached $R_{\rm inv} \le 2.0\%$ convergence within 4 node-hours (512 core-hours\footnote{Hereafter the figure is for the single computation, i.e. effectively twice longer in a few cases. We count physical cores and not hyperthreads. Later profiling also showed a possibility of $\approx 2\times$ improvement by using 64 threads and physical cores instead of 128.}) on the NERSC Perlmutter supercomputer.
The ELG covariance matrices reached $R_{\rm inv} \le 2.9\%$ convergence within 10 node-hours (1280 core-hours).
BGS reached $R_{\rm inv} < 11.5\%$ convergence within 12 node-hours (1536 core-hours).

The computations have become longer relative to \cite{RascalC-DESI-M2} whereas $R_{\rm inv}$ convergence figures have worsened (i.e. increased).
In this previous work the maximum $R_{\rm inv}$ was $0.63\%$.
The number of 4-point configurations contributing to the sums (\cref{eq:Cov2x2_234_Point_Defs_Legendre_Projected,eq:Cov2x2_234_Point_Defs-jack-legendre}) for LRG NGC or SGC in this work ($2.4\E{12}$) is very close to the analogous number for full LRG in \cite{RascalC-DESI-M2} ($2.5\E{12}$), where the sums (\cref{eq:Cov2x2_234_Point_Defs,eq:Cov2x2_234_Point_Defs-jack}) were evaluated for a single angular bin.

The processing of each 4-point configuration becomes longer in Legendre mode.
This can be expected because, as mentioned earlier, a given configuration contributes only to one pair of angular bins, but to all Legendre multipoles.
Other factors, like CPU differences, parallelism efficiency, and a larger number of 3- and 2-point configurations in this work, may also affect the runtimes.

The increase of the $R_{\rm inv}$ convergence measure can be primarily attributed to three times more observables in the covariance matrix.
In this work, we have the same number of radial bins as in \cite{RascalC-DESI-M2}, but three multipoles instead of only monopole.
A higher-dimensional covariance matrix requires more configurations sampled for the same relative precision.
Additionally, ELG and especially BGS are more challenging due to the increasing importance of the 4-point term relative to the 3- and 2-point terms.
The more points, the more configurations need to be sampled for the same precision of the term.
Still, the $R_{\rm inv}$ we obtained are significantly smaller than the expected deviation of the mock sample covariance estimate from the true covariance matrix ($R_{\rm inv} \approx 37\%$\footnote{As we show later in \cref{tab:cov-comparison-full}, this number is based on 1000 mock realizations. For a fixed number of samples, the relative precision of the sample covariance also worsens with the number of observables.}).

\subsection{Shot-noise rescaling values}
\label{sec:shot-noise-rescaling-values}

\begin{table}[tb]
    \centering
    \begin{tabular}{|c|c|c|c|c|}
\hline
\multirow{2}{*}{$\snrescaling$} & \multicolumn{2}{|c|}{NGC} & \multicolumn{2}{|c|}{SGC} \\
\cline{2-5}
 & Mocks & Data & Mocks & Data \\
\hline
LRG pre-recon & $0.743 \pm 0.012$ & 0.865 & $0.7935 \pm 0.0081$ & 0.953 \\
\hline
LRG post-recon & $0.770 \pm 0.010$ & 0.836 & $0.809 \pm 0.011$ & 0.969 \\
\hline
ELG pre-recon & $0.3757 \pm 0.0043$ & 0.687 & $0.4018 \pm 0.0014$ & 0.718 \\
\hline
ELG post-recon & $0.3789 \pm 0.0044$ & 0.693 & $0.4051 \pm 0.0044$ & 0.737 \\
\hline
BGS pre-recon & $0.792 \pm 0.012$ & 0.855 & $0.8198 \pm 0.0091$ & 0.897 \\
\hline
BGS post-recon & $0.812 \pm 0.012$ & 0.872 & $0.8447 \pm 0.0095$ & 0.934 \\
\hline
    \end{tabular}
    \caption{Shot-noise rescaling values $\snrescaling$ for mocks (10 realizations of FFA \ezmocks{}) and data. 
    }
    \label{tab:shot-noise-rescaling}
\end{table}

Our method uses a rescaling of the shot noise contribution to account for differences between the true small-scale contributions and our Gaussian approximation, as we pointed out in \cref{sec:cov-estimation-review}.
After reaching a relatively uniform convergence level in the previous section, we should investigate the values of the shot-noise rescaling parameter.

\cref{tab:shot-noise-rescaling} shows the shot-noise rescaling values obtained for mocks and DESI DR1 data according to the fiducial procedure (\cref{fig:pipeline-jack}).
Interestingly, we find that all the shot-noise rescaling values are smaller than 1;
i.e., the jackknife variations are smaller than predicted by Gaussian approximation with standard Poisson shot noise.
Previous mock studies \cite{rascal,rascal-jackknife,rascalC,RascalC-DESI-M2}, on the contrary, obtained shot-noise rescaling values greater than 1.

We also see that the shot-noise rescaling values are significantly lower for mocks than for the data.
The difference is most pronounced for ELGs, which have the lowest shot-noise rescaling values for both mocks and data.
ELGs are also impacted most by the fiber incompleteness effect due to their lower priority compared to other dark-time targets, LRG and QSO \cite{ELG.TS.Raichoor.2023}.
This suggests that the low shot-noise rescaling is related to fiber assignment.

To test whether fiber assignment is the main cause for low and different shot-noise rescaling values, we have performed additional \rascalc{} computations using more realistic \abacus{} mocks and the DESI fiber assignment algorithm (AMTL).
We only used one realization for LRG and ELG in each case to save computing time.

\begin{table}[tb!]
    \centering
    \begin{tabular}{|c|c|c|c|c|c|c|}
\hline
\multirow{2}{*}{$\snrescaling$} & \multirow{2}{*}{Data} & \multicolumn{2}{|c|}{AMTL (correct)} & \multicolumn{2}{|c|}{FFA (approximate)} & Complete \\
\cline{3-7}
 &  & \abacus{} & \ezmocks{} & \abacus{} & \ezmocks{} & \abacus{} \\
\hline
LRG NGC & 0.865 & 0.845 & 0.886 & 0.721 & 0.758 & 0.934 \\
\hline
LRG SGC & 0.953 & 0.954 & 0.961 & 0.781 & 0.784 & 0.962 \\
\hline
ELG NGC & 0.687 & 0.649 & 0.672 & 0.378 & 0.380 & 0.965 \\
\hline
ELG SGC & 0.718 & 0.707 & 0.744 & 0.405 & 0.403 & 0.968 \\
\hline
    \end{tabular}
    \caption{Shot-noise rescaling values $\alpha_{\rm SN}$ (before reconstruction) for the data and two different types of mocks (\cref{sec:mock-types}) with two different fiber assignment models (\cref{sec:fiber-assignment-models}).
    ``Complete'' designates mocks before fiber assignment.
    We use a single mock realization for each category.}
    \label{tab:shot-noise-rescaling-ffa-altmtl}
\end{table}

\cref{tab:shot-noise-rescaling-ffa-altmtl} shows the shot-noise rescaling values for data and different mocks (\abacus{} or \ezmocks{}) with different fiber assignment models (FFA or AMTL).
We also include an \abacus{} mock before fiber assignment (``complete'').
We see that the fiber assignment modeling method makes a bigger difference than the type of the mocks.
The approximate fast fiber assignment gives the lowest shot-noise rescaling values.
The DESI fiber assignment algorithm applied to the mocks (AMTL) gives $\alpha_{\rm SN}$ very similar to the data.
The complete \abacus{} mocks have larger shot-noise rescaling values than data and fiber-assigned mocks.
This confirms the fiber assignment as a key factor affecting the $\alpha_{\rm SN}$ parameter.

The discrepancy in shot-noise rescaling values also raises concerns about the quality of approximations in the FFA algorithm.
However, of all the mock types only \ezmocks{} processed with FFA are numerous enough for a precise sample covariance estimate.
Therefore we continue using them in the remainder of this paper.

\begin{table}[tb!]
    \centering
    \begin{tabular}{|c|c|c|c|c|}
\hline
\multirow{2}{*}{$\snrescaling$} & \multicolumn{2}{|c|}{NGC} & \multicolumn{2}{|c|}{SGC} \\
\cline{2-5}
 & Jackknife & Mock sample & Jackknife & Mock sample \\
\hline
LRG pre & $0.743 \pm 0.012$ & $0.7417 \pm 0.0038$ & $0.7935 \pm 0.0081$ & $0.7906 \pm 0.0045$ \\
\hline
LRG post & $0.770 \pm 0.010$ & $0.7446 \pm 0.0029$ & $0.809 \pm 0.011$ & $0.7945 \pm 0.0032$ \\
\hline
ELG pre & $0.3757 \pm 0.0043$ & $0.3757 \pm 0.0013$ & $0.4018 \pm 0.0014$ & $0.4077 \pm 0.0016$ \\
\hline
ELG post & $0.3789 \pm 0.0044$ & $0.3751 \pm 0.0013$ & $0.4051 \pm 0.0044$ & $0.4067 \pm 0.0017$ \\
\hline
BGS pre & $0.792 \pm 0.012$ & $0.7916 \pm 0.0068$ & $0.8198 \pm 0.0091$ & $0.827 \pm 0.014$ \\
\hline
BGS post & $0.812 \pm 0.012$ & $0.8148 \pm 0.0070$ & $0.8447 \pm 0.0095$ & $0.844 \pm 0.013$ \\
\hline
    \end{tabular}
    \caption{Shot-noise rescaling values for the single-mock runs calibrated on jackknife (\cref{fig:pipeline-jack}) and mock sample covariances (\cref{fig:pipeline-mocks}).}
    \label{tab:shot-noise-rescaling-jack-mocks}
\end{table}

We performed the final consistency tests of the shot-noise optimization procedure in light of the concerns about fiber assignment and jackknife discussed in \cref{sec:fiber-assignment-models}.
We optimized the shot-noise rescaling based on mock sample covariance (the process is shown schematically in \cref{fig:pipeline-mocks}).
We show the resulting values along with the baseline, jackknife-based ones in \cref{tab:shot-noise-rescaling-jack-mocks}, and find them very close for all cases.
In other words, calibration of shot-noise rescaling on jackknives still brings us close to an optimal fit on the mocks.
With that, we have decided to proceed further with the validation process using the jackknife-calibrated shot-noise rescaling values.
This will show how close this nearly optimal fit is to the mock sample covariance.

\section{Comparison between semi-analytical and mock sample covariance matrices}
\label{sec:cov-comparison-results}

Thus we reach the last validation part: comparison of the semi-analytical covariances obtained from single mock catalogs with the mock sample covariance matrices.
We compute three similarity measures (\cref{sec:cov-comparison-methods}) between these covariance matrices for each tracer before and after standard BAO reconstruction (pre- and post-recon).
Instead of presenting 10 numbers (corresponding to each \rascalc{} realization) for each case in the paper, we provide their mean and standard deviation.
The full set of comparison measure values is available in the supplementary material\footnote{\supplementarylink{}}.
We also provide the ``perfect'' reference row, listing the statistical properties of the similarity measures between the true covariance and a mock sample covariance matrix based on their dimension (also obtained in \cref{sec:cov-comparison-methods}).

We note that the standard deviation in the ``perfect'' row is different and independent from the others.
The ``perfect'' standard deviation characterizes the distribution of the random difference between the sample covariance estimate from 1000 realizations and the true covariance.
In each other row (tracer + pre- or post-recon combination) the sample covariance estimate is fixed, and the standard deviation describes the scatter resulting from 10 different \rascalc{} covariance realizations.

Ideally, every \rascalc{} covariance matrix would be consistent with the true covariance.
To see whether this is the case, we compute the difference between each comparison measure for each \rascalc{} covariance realization and the corresponding ``perfect'' mean in ``perfect'' standard deviations.
We summarize the 10 resulting quantities by the mean and standard deviation as well and provide them below the summary of the similarity measure itself.
This sets the common structure for all tables in this section (i.e., \cref{tab:cov-comparison-full,tab:cov-comparison-shape-range,tab:cov-comparison-BAO-range,tab:cov-comparison-BAO-parameters,tab:cov-comparison-ShapeFit-parameters,tab:cov-comparison-direct-fit-parameters}.).

\subsection{Observables}
\label{sec:cov-comparison-obs}

We begin with the covariances for correlation function multipoles, in other words, in the space of observables. 

\begin{table}[tb]
\centering
\begin{tabular}{|c|c|c|c|}
\hline
 & $D_{\rm KL} ({\bf C}_R^{-1}, {\bf C}_S)$ & $R_{\rm inv} ({\bf C}_R^{-1}, {\bf C}_S)$ & $\chi^2_{\rm red} ({\bf C}_R^{-1}, {\bf C}_S)$ \\
\hline
Perfect & $4.817 \pm 0.070$ & $0.3690 \pm 0.0031$ & $1.0000 \pm 0.0039$ \\
\hline
\multirow{2}{*}{LRG pre-recon} & $4.856 \pm 0.029$ & $0.3678 \pm 0.0049$ & $0.989 \pm 0.016$ \\
 & ($0.55 \pm 0.41$)$\sigma$ & ($-0.4 \pm 1.6$)$\sigma$ & ($-2.7 \pm 4.1$)$\sigma$ \\
\hline
\multirow{2}{*}{LRG post-recon} & $4.977 \pm 0.050$ & $0.3581 \pm 0.0034$ & $0.957 \pm 0.014$ \\
 & ($2.27 \pm 0.70$)$\sigma$ & ($-3.5 \pm 1.1$)$\sigma$ & ($-11.3 \pm 3.6$)$\sigma$ \\
\hline
\multirow{2}{*}{ELG pre-recon} & $4.811 \pm 0.024$ & $0.3700 \pm 0.0055$ & $1.000 \pm 0.015$ \\
 & ($-0.09 \pm 0.34$)$\sigma$ & ($0.3 \pm 1.8$)$\sigma$ & ($0.1 \pm 3.8$)$\sigma$ \\
\hline
\multirow{2}{*}{ELG post-recon} & $5.001 \pm 0.018$ & $0.3701 \pm 0.0042$ & $0.986 \pm 0.012$ \\
 & ($2.61 \pm 0.26$)$\sigma$ & ($0.4 \pm 1.4$)$\sigma$ & ($-3.6 \pm 3.1$)$\sigma$ \\
\hline
\multirow{2}{*}{BGS pre-recon} & $5.129 \pm 0.052$ & $0.3824 \pm 0.0081$ & $0.997 \pm 0.016$ \\
 & ($4.43 \pm 0.73$)$\sigma$ & ($4.4 \pm 2.6$)$\sigma$ & ($-0.8 \pm 4.2$)$\sigma$ \\
\hline
\multirow{2}{*}{BGS post-recon} & $5.177 \pm 0.077$ & $0.3810 \pm 0.0079$ & $0.994 \pm 0.016$ \\
 & ($5.1 \pm 1.1$)$\sigma$ & ($3.9 \pm 2.6$)$\sigma$ & ($-1.4 \pm 4.1$)$\sigma$ \\
\hline
\end{tabular}
\caption{Summary of full observable-space comparison of \rascalc{} covariances with the sample covariances (135 bins, $s=20-200~\ihMpc$, monopole, quadrupole and hexadecapole).}
\label{tab:cov-comparison-full}
\end{table}

\cref{tab:cov-comparison-full} shows the comparison measures for the full covariance matrices, covering $s=20-200~\ihMpc$ radial bins for all three multipoles.
We can see that some \rascalc{} results deviate significantly from the ``perfect'' (i.e., the true covariance behavior).
We remind that the KL divergence and $R_{\rm inv}$ accumulate deviations in all ``directions''.
The KL divergences exceed the ideal expectation value by nearly $3\sigma$ for LRG and ELG post-recon, whereas for BGS they are even further from perfect.
$R_{\rm inv}$ are high with a larger scatter for BGS.
In the reduced chi-squared, which captures the overall ``scaling'' with higher accuracy, the mean values for the \rascalc{} runs are shifted significantly for LRG and ELG post-recon, and the mock-to-mock scatter is high in all the cases.

\begin{table}[tb]
\centering
\begin{tabular}{|c|c|c|c|}
\hline
 & $D_{\rm KL} ({\bf C}_R^{-1}, {\bf C}_S)$ & $R_{\rm inv} ({\bf C}_R^{-1}, {\bf C}_S)$ & $\chi^2_{\rm red} ({\bf C}_R^{-1}, {\bf C}_S)$ \\
\hline
Perfect & $2.260 \pm 0.049$ & $0.3067 \pm 0.0036$ & $1.0000 \pm 0.0046$ \\
\hline
\multirow{2}{*}{LRG pre-recon} & $2.307 \pm 0.022$ & $0.3056 \pm 0.0037$ & $0.983 \pm 0.016$ \\
 & ($0.99 \pm 0.46$)$\sigma$ & ($-0.3 \pm 1.0$)$\sigma$ & ($-3.7 \pm 3.4$)$\sigma$ \\
\hline
\multirow{2}{*}{LRG post-recon} & $2.333 \pm 0.027$ & $0.2998 \pm 0.0024$ & $0.960 \pm 0.014$ \\
 & ($1.52 \pm 0.56$)$\sigma$ & ($-1.93 \pm 0.68$)$\sigma$ & ($-8.7 \pm 3.0$)$\sigma$ \\
\hline
\multirow{2}{*}{ELG pre-recon} & $2.2578 \pm 0.0095$ & $0.3050 \pm 0.0044$ & $0.995 \pm 0.014$ \\
 & ($-0.04 \pm 0.20$)$\sigma$ & ($-0.5 \pm 1.2$)$\sigma$ & ($-1.0 \pm 3.1$)$\sigma$ \\
\hline
\multirow{2}{*}{ELG post-recon} & $2.292 \pm 0.013$ & $0.3044 \pm 0.0033$ & $0.987 \pm 0.012$ \\
 & ($0.67 \pm 0.28$)$\sigma$ & ($-0.65 \pm 0.93$)$\sigma$ & ($-2.8 \pm 2.6$)$\sigma$ \\
\hline
\multirow{2}{*}{BGS pre-recon} & $2.414 \pm 0.025$ & $0.3140 \pm 0.0066$ & $0.987 \pm 0.016$ \\
 & ($3.18 \pm 0.52$)$\sigma$ & ($2.0 \pm 1.9$)$\sigma$ & ($-2.7 \pm 3.4$)$\sigma$ \\
\hline
\multirow{2}{*}{BGS post-recon} & $2.479 \pm 0.038$ & $0.3202 \pm 0.0065$ & $0.993 \pm 0.016$ \\
 & ($4.52 \pm 0.78$)$\sigma$ & ($3.8 \pm 1.8$)$\sigma$ & ($-1.5 \pm 3.4$)$\sigma$ \\
\hline
\end{tabular}
\caption{Summary of observable-space comparison of \rascalc{} covariances with the sample covariances restricted to the range of ShapeFit and full modeling fit (93 bins, $s=28-152~\ihMpc$, monopole, quadrupole and hexadecapole).}
\label{tab:cov-comparison-shape-range}
\end{table}

We continue the comparisons in \cref{tab:cov-comparison-shape-range}, now cutting the range to $s=28-152~\ihMpc$ as is common for full-shape fits (ShapeFit and direct).
The KL divergences and $R_{\rm inv}$ become more consistent with the perfect reference cases for LRG and ELG, but remain high for BGS.
The scaling difference (in reduced chi-squared) remains similar.

\begin{table}[tb]
\centering
\begin{tabular}{|c|c|c|c|}
\hline
 & $D_{\rm KL} ({\bf C}_R^{-1}, {\bf C}_S)$ & $R_{\rm inv} ({\bf C}_R^{-1}, {\bf C}_S)$ & $\chi^2_{\rm red} ({\bf C}_R^{-1}, {\bf C}_S)$ \\
\hline
Perfect & $0.702 \pm 0.027$ & $0.2303 \pm 0.0046$ & $1.0000 \pm 0.0062$ \\
\hline
\multirow{2}{*}{LRG pre-recon} & $0.763 \pm 0.014$ & $0.2351 \pm 0.0023$ & $0.982 \pm 0.016$ \\
 & ($2.29 \pm 0.54$)$\sigma$ & ($1.04 \pm 0.49$)$\sigma$ & ($-2.9 \pm 2.6$)$\sigma$ \\
\hline
\multirow{2}{*}{LRG post-recon} & $0.732 \pm 0.013$ & $0.2281 \pm 0.0019$ & $0.964 \pm 0.015$ \\
 & ($1.14 \pm 0.50$)$\sigma$ & ($-0.48 \pm 0.40$)$\sigma$ & ($-5.7 \pm 2.3$)$\sigma$ \\
\hline
\multirow{2}{*}{ELG pre-recon} & $0.7195 \pm 0.0083$ & $0.2317 \pm 0.0039$ & $0.999 \pm 0.015$ \\
 & ($0.65 \pm 0.31$)$\sigma$ & ($0.30 \pm 0.86$)$\sigma$ & ($-0.2 \pm 2.4$)$\sigma$ \\
\hline
\multirow{2}{*}{ELG post-recon} & $0.6903 \pm 0.0088$ & $0.2278 \pm 0.0029$ & $0.995 \pm 0.012$ \\
 & ($-0.45 \pm 0.33$)$\sigma$ & ($-0.54 \pm 0.62$)$\sigma$ & ($-0.8 \pm 1.9$)$\sigma$ \\
\hline
\multirow{2}{*}{BGS pre-recon} & $0.796 \pm 0.021$ & $0.2419 \pm 0.0079$ & $0.982 \pm 0.018$ \\
 & ($3.54 \pm 0.78$)$\sigma$ & ($2.5 \pm 1.7$)$\sigma$ & ($-2.8 \pm 2.8$)$\sigma$ \\
\hline
\multirow{2}{*}{BGS post-recon} & $0.777 \pm 0.031$ & $0.2462 \pm 0.0090$ & $1.011 \pm 0.017$ \\
 & ($2.8 \pm 1.2$)$\sigma$ & ($3.4 \pm 1.9$)$\sigma$ & ($1.8 \pm 2.7$)$\sigma$ \\
\hline
\end{tabular}
\caption{Summary of observable-space comparison of \rascalc{} covariances with the sample covariances restricted to the range of BAO fits (52 bins, $s=48-152~\ihMpc$, monopole and quadrupole).}
\label{tab:cov-comparison-BAO-range}
\end{table}

We perform the final set of observable-space comparisons in \cref{tab:cov-comparison-BAO-range}, further restricting the range to $s=48-152~\ihMpc$ and using only monopole and quadrupole, without hexadecapole.
We do not see significant consistency changes from the previous case.

We note that the abovementioned differences are relatively small.
In the reduced chi-squared, they are at most $(4.3 \pm 1.4)\%$ for LRG post-recon in the widest range, and in $R_{\rm inv}$ -- no more than a percent or two on top of $23-37\%$ caused by the finite sample size.
We should ask whether we trust the realism of the mocks to such a high level in all aspects of the correlation function multipoles.
Moreover, for the real survey matching the clustering between data and simulations will become an additional issue for mocks.
On the other hand, our \rascalc{} computations use the correlation function measured directly from single mock catalogs similar to real data.
Thus we conclude that the semi-analytic method performance is very compelling.

\subsection{Model parameters}
\label{sec:cov-comparison-param}

We proceed to project the covariance matrices (as described in \cref{sec:param-space-fisher}) to the parameters of the models listed in \cref{sec:theoretical-models-fits}.

We use the same model Jacobian (\cref{eq:model-jacobian}) for projecting \rascalc{} covariances (\cref{eq:parameter-covariance-rascalc}) and the mock sample covariances (\cref{eq:parameter-covariance-sample}) into the parameter space.
We compute the partial derivatives at the best fit of each model to the mean clustering of all the available mocks for each galaxy type.
Since BAO reconstruction changes clustering, we compute separate Jacobians before and after reconstruction.
We use the mock sample covariance matrix in this fit.
Using different Jacobian matrices for each covariance matrix would complicate the comparison.

A more thorough investigation of the covariance matrix impact on the model fits is presented in the companion paper \cite{KP4s6-Forero-Sanchez}.
They perform fits to each mock clustering using different covariance matrices (mock sample and semi-analytical) and compare the resulting best parameter values as well as errorbar estimates.
On the flip side, with such a detailed approach \cite{KP4s6-Forero-Sanchez} can not test multiple \rascalc{} single-mock covariances.

\subsubsection{BAO}
\label{sec:cov-comparison-bao-param}

\begin{table}[tb]
\centering
\begin{tabular}{|c|c|c|c|}
\hline
 & $D_{\rm KL} ({\bf C}_R^{-1}, {\bf C}_S)$ & $R_{\rm inv} ({\bf C}_R^{-1}, {\bf C}_S)$ & $\chi^2_{\rm red} ({\bf C}_R^{-1}, {\bf C}_S)$ \\
\hline
Perfect & $0.0015 \pm 0.0012$ & $0.051 \pm 0.021$ & $1.000 \pm 0.032$ \\
\hline
\multirow{2}{*}{LRG pre-recon} & $0.00103 \pm 0.00066$ & $0.042 \pm 0.014$ & $0.960 \pm 0.014$ \\
 & ($-0.39 \pm 0.53$)$\sigma$ & ($-0.39 \pm 0.67$)$\sigma$ & ($-1.27 \pm 0.45$)$\sigma$ \\
\hline
\multirow{2}{*}{LRG post-recon} & $0.00110 \pm 0.00026$ & $0.0458 \pm 0.0051$ & $0.9734 \pm 0.0098$ \\
 & ($-0.32 \pm 0.21$)$\sigma$ & ($-0.23 \pm 0.24$)$\sigma$ & ($-0.84 \pm 0.31$)$\sigma$ \\
\hline
\multirow{2}{*}{ELG pre-recon} & $0.00069 \pm 0.00036$ & $0.0367 \pm 0.0093$ & $1.033 \pm 0.010$ \\
 & ($-0.66 \pm 0.29$)$\sigma$ & ($-0.65 \pm 0.44$)$\sigma$ & ($1.04 \pm 0.32$)$\sigma$ \\
\hline
\multirow{2}{*}{ELG post-recon} & $0.00078 \pm 0.00045$ & $0.037 \pm 0.012$ & $0.967 \pm 0.012$ \\
 & ($-0.59 \pm 0.37$)$\sigma$ & ($-0.63 \pm 0.55$)$\sigma$ & ($-1.05 \pm 0.38$)$\sigma$ \\
\hline
\multirow{2}{*}{BGS pre-recon} & $0.0042 \pm 0.0016$ & $0.088 \pm 0.015$ & $0.915 \pm 0.014$ \\
 & ($2.2 \pm 1.3$)$\sigma$ & ($1.74 \pm 0.71$)$\sigma$ & ($-2.68 \pm 0.46$)$\sigma$ \\
\hline
\multirow{2}{*}{BGS post-recon} & $0.00027 \pm 0.00024$ & $0.0213 \pm 0.0093$ & $0.992 \pm 0.013$ \\
 & ($-1.00 \pm 0.19$)$\sigma$ & ($-1.38 \pm 0.44$)$\sigma$ & ($-0.24 \pm 0.40$)$\sigma$ \\
\hline
\end{tabular}
\caption{Summary of parameter-space comparison of \rascalc{} covariances with the sample covariances projected to the BAO fit parameters, $\alphaiso$ and $\alphaap$.}
\label{tab:cov-comparison-BAO-parameters}
\end{table}

In \cref{tab:cov-comparison-BAO-parameters} we compare the covariances projected to the BAO parameters, the scaling parameters $\alphaiso$ and $\alphaap$\footnote{Because we use Fisher matrix formalism, the covariance matrices projected for $\alpha_\parallel$ and $\alpha_\perp$ should have the same comparison measures.}.
The comparison measures look consistent with the perfect reference case.
The most significant deviations are seen in BGS pre-recon: both the KL divergence and $R_{\rm inv}$ are high, while the reduced chi-squared is almost 3 sigma low on average (meaning that the \rascalc{} covariance is ``larger'' than the mock sample one).
The mock-to-mock scatter in \rascalc{} results is always lower than the noise expected from the finite mock sample size.

\begin{figure}[tb]
\centering
\includegraphics[width=\textwidth]{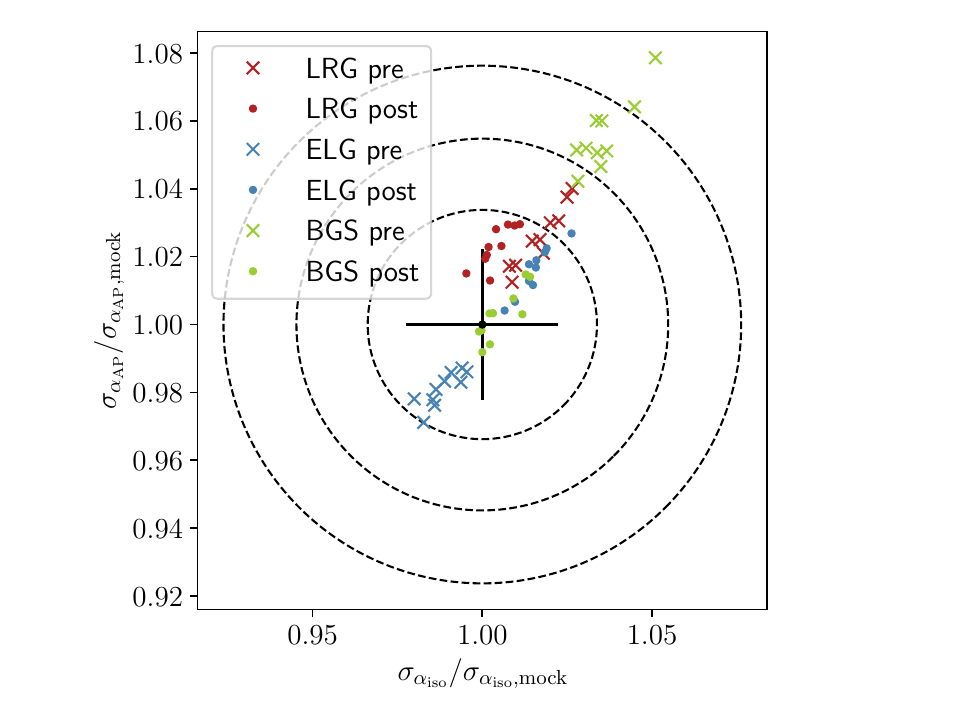}
\caption{Comparison of the projected errorbars for the BAO scale parameters, normalized to the values obtained from the mock sample covariance.
The cross shows one-dimensional relative precisions ($\qty(2\qty(n_S-1))^{-1/2} \approx 2.2\%$ \cite{RascalC-DESI-M2}) following from the \ezmocks{} sample size ($n_S=1000$), and the dashed ellipses approximately correspond to two-dimensional 68\%, 95\% and 99.7\% confidence regions in this 2D space of errorbars.
Here the correlation of errorbars is ignored; it varies in different cases but is too small to notice ($\lesssim 0.04$).}
\label{fig:BAO-errorbars}
\end{figure}

We have also plotted the errorbars on $\alphaiso$ and $\alphaap$ against each other in \cref{fig:BAO-errorbars}.
They corroborate \cref{tab:cov-comparison-BAO-parameters}.
Almost all deviations are within the 99.7\% contour.
The only exceptions are two BGS pre-recon realizations.
However, we note that these plots do not show the covariance of the parameters, which is taken into account with KL divergence and $R_{\rm inv}$.

\subsubsection{Full shape: ShapeFit and full modeling}
\label{sec:cov-comparison-fullshape-param}

\begin{table}[tb]
\centering
\begin{tabular}{|c|c|c|c|}
\hline
 & $D_{\rm KL} ({\bf C}_R^{-1}, {\bf C}_S)$ & $R_{\rm inv} ({\bf C}_R^{-1}, {\bf C}_S)$ & $\chi^2_{\rm red} ({\bf C}_R^{-1}, {\bf C}_S)$ \\
\hline
Perfect & $0.0050 \pm 0.0023$ & $0.069 \pm 0.016$ & $1.000 \pm 0.022$ \\
\hline
\multirow{2}{*}{LRG pre-recon} & $0.0059 \pm 0.0023$ & $0.073 \pm 0.013$ & $0.969 \pm 0.020$ \\
 & ($0.4 \pm 1.0$)$\sigma$ & ($0.25 \pm 0.81$)$\sigma$ & ($-1.39 \pm 0.89$)$\sigma$ \\
\hline
\multirow{2}{*}{LRG post-recon} & $0.00739 \pm 0.00087$ & $0.0840 \pm 0.0052$ & $0.984 \pm 0.022$ \\
 & ($1.06 \pm 0.39$)$\sigma$ & ($0.95 \pm 0.33$)$\sigma$ & ($-0.73 \pm 0.98$)$\sigma$ \\
\hline
\multirow{2}{*}{ELG pre-recon} & $0.0063 \pm 0.0029$ & $0.078 \pm 0.015$ & $1.005 \pm 0.027$ \\
 & ($0.6 \pm 1.3$)$\sigma$ & ($0.59 \pm 0.94$)$\sigma$ & ($0.2 \pm 1.2$)$\sigma$ \\
\hline
\multirow{2}{*}{ELG post-recon} & $0.0029 \pm 0.0013$ & $0.052 \pm 0.011$ & $0.999 \pm 0.017$ \\
 & ($-0.95 \pm 0.58$)$\sigma$ & ($-1.06 \pm 0.68$)$\sigma$ & ($-0.06 \pm 0.78$)$\sigma$ \\
\hline
\multirow{2}{*}{BGS pre-recon} & $0.0071 \pm 0.0030$ & $0.083 \pm 0.019$ & $0.996 \pm 0.031$ \\
 & ($0.9 \pm 1.3$)$\sigma$ & ($0.9 \pm 1.2$)$\sigma$ & ($-0.2 \pm 1.4$)$\sigma$ \\
\hline
\multirow{2}{*}{BGS post-recon} & $0.0067 \pm 0.0027$ & $0.080 \pm 0.017$ & $0.992 \pm 0.030$ \\
 & ($0.7 \pm 1.2$)$\sigma$ & ($0.7 \pm 1.1$)$\sigma$ & ($-0.3 \pm 1.3$)$\sigma$ \\
\hline
\end{tabular}
\caption{Summary of parameter-space comparison of \rascalc{} covariances with the sample covariances projected to the ShapeFit parameters: $\alphaiso$, $\alphaap$, $dm$ and $df$.}
\label{tab:cov-comparison-ShapeFit-parameters}
\end{table}

\cref{tab:cov-comparison-ShapeFit-parameters} shows the comparison measures for the covariances projected to the ShapeFit parameters: $\alphaiso$, $\alphaap$, $dm$ and $df$.
We do not see significant statistical deviations from the perfect case.
This is the only case when BGS (both pre- and post-recon) are not particularly far from the reference.
The mock-to-mock scatter in \rascalc{} results is smaller than or comparable with the noise in the sample covariances.

\begin{figure}[tb]
\centering
\includegraphics[width=\textwidth]{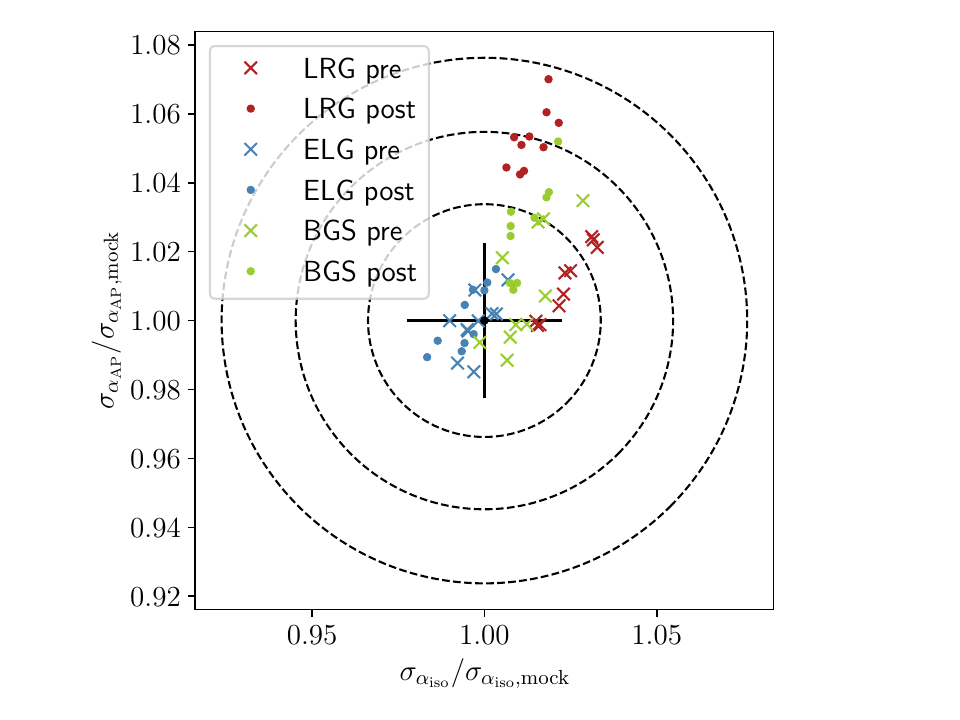}
\caption{Same as \cref{fig:BAO-errorbars} but with errorbars for the scale parameters following from ShapeFit.}
\label{fig:shapefit-scale-errorbars}
\end{figure}

Additionally, in \cref{fig:shapefit-scale-errorbars} we compare the projected errorbars for the scaling parameters, $\alphaiso$ and $\alphaap$.
All the points fall within the 99.7\% confidence region in this 2D space, although the other parameters and cross-correlations are ignored in this picture, unlike in \cref{tab:cov-comparison-ShapeFit-parameters}.

\begin{table}[tb]
\centering
\begin{tabular}{|c|c|c|c|}
\hline
 & $D_{\rm KL} ({\bf C}_R^{-1}, {\bf C}_S)$ & $R_{\rm inv} ({\bf C}_R^{-1}, {\bf C}_S)$ & $\chi^2_{\rm red} ({\bf C}_R^{-1}, {\bf C}_S)$ \\
\hline
Perfect & $0.0050 \pm 0.0023$ & $0.069 \pm 0.016$ & $1.000 \pm 0.022$ \\
\hline
\multirow{2}{*}{LRG pre-recon} & $0.0099 \pm 0.0034$ & $0.093 \pm 0.014$ & $0.951 \pm 0.020$ \\
 & ($2.2 \pm 1.5$)$\sigma$ & ($1.50 \pm 0.89$)$\sigma$ & ($-2.20 \pm 0.91$)$\sigma$ \\
\hline
\multirow{2}{*}{LRG post-recon} & $0.00360 \pm 0.00068$ & $0.0584 \pm 0.0055$ & $0.976 \pm 0.016$ \\
 & ($-0.63 \pm 0.30$)$\sigma$ & ($-0.66 \pm 0.35$)$\sigma$ & ($-1.08 \pm 0.71$)$\sigma$ \\
\hline
\multirow{2}{*}{ELG pre-recon} & $0.0061 \pm 0.0013$ & $0.0783 \pm 0.0094$ & $1.013 \pm 0.024$ \\
 & ($0.48 \pm 0.60$)$\sigma$ & ($0.59 \pm 0.59$)$\sigma$ & ($0.6 \pm 1.1$)$\sigma$ \\
\hline
\multirow{2}{*}{ELG post-recon} & $0.0033 \pm 0.0014$ & $0.055 \pm 0.011$ & $0.985 \pm 0.016$ \\
 & ($-0.75 \pm 0.60$)$\sigma$ & ($-0.85 \pm 0.72$)$\sigma$ & ($-0.68 \pm 0.72$)$\sigma$ \\
\hline
\multirow{2}{*}{BGS pre-recon} & $0.0101 \pm 0.0073$ & $0.099 \pm 0.040$ & $1.013 \pm 0.035$ \\
 & ($2.3 \pm 3.2$)$\sigma$ & ($1.9 \pm 2.5$)$\sigma$ & ($0.6 \pm 1.5$)$\sigma$ \\
\hline
\multirow{2}{*}{BGS post-recon} & $0.0083 \pm 0.0045$ & $0.090 \pm 0.027$ & $1.003 \pm 0.035$ \\
 & ($1.5 \pm 2.0$)$\sigma$ & ($1.3 \pm 1.7$)$\sigma$ & ($0.1 \pm 1.5$)$\sigma$ \\
\hline
\end{tabular}
\caption{Summary of parameter-space comparison of \rascalc{} covariances with the sample covariances projected to the full modeling parameters: $h$, $\omega_{\rm cdm}$, $\omega_b$ and $\log A_s$.}
\label{tab:cov-comparison-direct-fit-parameters}
\end{table}

Finally, in \cref{tab:cov-comparison-direct-fit-parameters} we provide the comparison results for the covariances projected to the full modeling parameters, $h$, $\omega_{\rm cdm}$, $\omega_b$ and $\log A_s$.
We see deviations exceeding 3 sigma in most of the measures for LRG pre-recon and of BGS; LRG post-recon and ELG look consistent.
Still, the differences we see are at a few percent level, and may be caused by limited realism of the mocks.

\section{Conclusions}
\label{sec:conclusion}

We present and validate the DESI DR1 pipeline for the semi-analytical covariance matrices of the galaxy 2-point correlation functions on realistic mock catalogs including a model of fiber assignment.

We develop a streamlined procedure for the estimation of the semi-analytical covariance matrices for Legendre moments of the 2PCF in separation bins with the \rascalc{} code \cite{rascalC}.
The previous implementation \cite{rascalC-legendre-3} required an additional computation with angular bins to mimic the non-Gaussian effects by calibrating the shot-noise rescaling value on the jackknife covariance matrix estimate.
Now we can perform this calibration together with the construction of the covariance model for Legendre moments.
This allowed for more efficient massive production of covariance matrices for all the tracers, redshift bins, and galactic caps of DESI DR1 galaxies and quasars data.

We apply this updated pipeline to mock catalogs with fast (approximate) fiber assignment, representative of DESI DR1.
We cover 3 selected tracers and redshift bins ({\tt LRG3}, {\tt ELG2} and {\tt BGS} according to \cite{DESI2024.III.KP4}), without and with standard BAO reconstruction applied.
We use a single mock catalog and repeat the procedure for 10 different realizations to assess the impact of realistic fluctuations in the input quantities.

First, we note the difference between the shot-noise rescaling values obtained for the mocks with fast fiber assignment and the data in \cref{sec:shot-noise-rescaling-values}.
We show that the fiber assignment modeling (and the associated weighting scheme) is the main factor causing the difference.
Then we find the parameter calibration on jackknife and mock sample covariance yields very close results in the mock runs.
This motivates us to proceed with the comparison of the covariance matrices.

Then, we apply the set of compact measures of covariance matrix similarity from \cite{RascalC-DESI-M2}.
We use covariances in the observable space (correlation function multipoles) in \cref{sec:cov-comparison-obs}, as well as projected linearly (through Fisher matrix formalism) to the parameters of different models in \cref{sec:cov-comparison-param}.
We find some deviations that cannot be explained solely by the finite sample size limiting the accuracy of the mock-based covariance.
However, these differences are at a few percent level.

We argue that this level of agreement is sufficient for real-world applications.
First, mocks necessarily involve approximations to make sufficiently many catalogs.
Second, simulations are never perfectly representative of data in terms of clustering due to the limited precision of the measurements.
Blinding can aggravate this issue by forcing the mock tuning to rely on earlier, smaller samples with larger uncertainties in their clustering.
Additionally, some simulations, including \ezmocks{} \cite{DESI2024.III.KP4,KP3s8-Zhao}, are only matched to the 1- and 2-point statistics of the data, whereas the covariance matrices are impacted by 3- and 4-point functions.
Therefore matching the mocks perfectly is not necessarily a reasonable goal.

Our results for the BAO model (\cref{sec:cov-comparison-bao-param}) are particularly important because \rascalc{} covariance matrices are used in the DESI DR1 baryon acoustic oscillations analysis \cite{DESI2024.III.KP4}.
The errorbars of the scale parameters ($\alphaiso$ and $\alphaap$) predicted from \rascalc{} agree to $\le 8\%$ with the mock sample covariance.
The standard deviation expected from the sample covariance of 1000 mocks itself is $\approx 2.4\%$.
When we exclude BGS before reconstruction (not used for the main DESI DR1 BAO), the errorbars agree within $\approx 5\%$.
Therefore we report a close match between the semi-analytic and mock sample covariance.

We find the covariance matrices for the BGS sample less consistent in most comparisons\footnote{A notable exception is ShapeFit parameters in \cref{sec:cov-comparison-fullshape-param}.}.
We expected them to be more challenging for \rascalc{} because of higher number density and thus higher significance of the 4-point term compared to the 3- and 2-point terms.
This already caused slower convergence and could further demonstrate the limitations of the shot-noise rescaling.
However, BGS have been challenging for the \ezmocks{} as well \cite{KP3s8-Zhao}\footnote{Moreover, DESI BGS \ezmocks{} had not been used in \cite{BAO.EDR.Moon.2023}.}, so their sample covariance is likely to be a less robust reference than for LRG and ELG.

The observable-space results (\cref{sec:cov-comparison-obs}) may leave an impression that \rascalc{} performance worsened with fiber assignment\footnote{Or due to including not only monopole, but also quadrupole and hexadecapole.}.
However, we think the change of interpretation primarily causes this.
The previous \rascalc{} validation for early DESI data \cite{RascalC-DESI-M2} used an earlier version of \ezmocks{} without any fiber assignment effects.
The covariance matrices there also showed statistically significant variations from the sample covariance in observable space.
They were deemed acceptable as comparable to the scatter in semi-analytical results.
In this work we use a stricter interpretation, testing whether every \rascalc{} single-mock result is consistent with the perfect-case reference.
In future work, we could repeat the tests on the mocks before fiber assignment with all the tracers (and multipoles) used in this paper.

Another direction for further improvement is to study the dependence of shot-noise rescaling on fiber assignment incompleteness.
This could be achieved by running \rascalc{} on survey sub-regions with different completeness patterns, set by the number of passes.
It is also possible that a prescription for higher-point non-Gaussian correlations with a low number of parameters would give a better consistency with the reference than rescaling the covariance matrix terms in which they are nulled.
However, for such precision studies, it is instructive to have extremely reliable, realistic and numerous mocks.

The semi-analytic approach can extend beyond the standard 2-point function.
First, the cross-covariances of different tracers are provided in \cref{sec:cov-estimation-extra}.
The full cross-covariance has more bins and requires more mocks than a single covariance for validation at similar precision.
Consistent simulated catalogs for different tracers are also crucial for capturing realistic cross-correlations.
Second, \cite{rascalc-power-spectrum} have introduced the covariance matrices for the modified power spectra, including the cross-covariance with correlation functions.
Third, \cite{rascalC-legendre-3} have derived the covariance matrices for isotropic 3-point correlation functions.
It involves higher-order correlation functions up to 6 points.
Approximating all of them in fast mocks is challenging.
Thus these extensions require extra work, particularly on the mock side for validation.

In summary, we have confirmed \rascalc{} semi-analytical covariance matrices for 2PCF as a very viable alternative to the mock-based ones.
Despite the increase in the runtime of the \rascalc{} code (from 100-300 \cite{RascalC-DESI-M2} to 500-1500\footnote{See \cref{sec:runtime-intrinsic-convergence}. Potentially $\approx 2$ times smaller using the later profiling results.} core-hours), the method is far faster than calibrating, generating, and processing a suite of mocks numerous enough to give an adequate covariance matrix precision.
We have shown that the two methods produce similar results given the requirements of the DESI 2024 BAO analysis.
The speed advantage of the semi-analytic method permits easier exploration of situations where one cannot afford to regenerate mock catalogs, such as different assumptions about cosmology, galaxy-halo connection, or non-standard sample selections.
We therefore expect that such semi-analytic methods can be of broad utility for computing large-scale covariance matrices in wide-field surveys.

\section*{Acknowledgements}

MR and DJE are supported by U.S. Department of Energy grant DE-SC0013718 and as a Simons Foundation Investigator.
DFS acknowledges support from the Swiss National Science Foundation (SNF) "Cosmology with 3D Maps of the Universe" research grant, 200020\_175751 and 200020\_207379.
H-JS acknowledges support from the U.S. Department of Energy, Office of Science, Office of High Energy Physics under grant No. DE-SC0023241. H-JS also acknowledges support from Lawrence Berkeley National Laboratory and the Director, Office of Science, Office of High Energy Physics of the U.S. Department of Energy under Contract No. DE-AC02-05CH1123 during the sabbatical visit.

This material is based upon work supported by the U.S. Department of Energy (DOE), Office of Science, Office of High-Energy Physics, under Contract No. DE–AC02–05CH11231, and by the National Energy Research Scientific Computing Center, a DOE Office of Science User Facility under the same contract. Additional support for DESI was provided by the U.S. National Science Foundation (NSF), Division of Astronomical Sciences under Contract No. AST-0950945 to the NSF National Optical-Infrared Astronomy Research Laboratory; the Science and Technology Facilities Council of the United Kingdom; the Gordon and Betty Moore Foundation; the Heising-Simons Foundation; the French Alternative Energies and Atomic Energy Commission (CEA); the National Council of Humanities, Science and Technology of Mexico (CONAHCYT); the Ministry of Science and Innovation of Spain (MICINN), and by the DESI Member Institutions: \url{https://www.desi.lbl.gov/collaborating-institutions}. Any opinions, findings, and conclusions or recommendations expressed in this material are those of the author(s) and do not necessarily reflect the views of the U. S. National Science Foundation, the U. S. Department of Energy, or any of the listed funding agencies.

The authors are honored to be permitted to conduct scientific research on Iolkam Du’ag (Kitt Peak), a mountain with particular significance to the Tohono O’odham Nation.

This work has used the following software packages: \textsc{astropy} \citep{astropy:2013, astropy:2018, astropy:2022}, \textsc{Jupyter} \citep{2007CSE.....9c..21P, kluyver2016jupyter}, \textsc{matplotlib} \citep{Hunter:2007}, \textsc{numpy} \citep{numpy}, \pycorr{} \citep{pycorr,corrfunc-1,corrfunc-2}, \textsc{python} \citep{python}, \textsc{scipy} \citep{2020SciPy-NMeth, scipy_8092679}, and \textsc{scikit-learn} \citep{scikit-learn, sklearn_api, scikit-learn_10034229}.

This research has made use of NASA's Astrophysics Data System.
Software citation information has been aggregated using \texttt{\href{https://www.tomwagg.com/software-citation-station/}{The Software Citation Station}} \citep{software-citation-station-paper, software-citation-station-zenodo}.

\section*{Data availability}

The data used in this analysis will be made public along with the DESI Data Release 1 (details in \url{https://data.desi.lbl.gov/doc/releases/}).
The code used in the paper is publicly available at \url{https://github.com/oliverphilcox/RascalC} and \url{https://github.com/cosmodesi/RascalC-scripts}.
The data from the plots, the full set of computed comparison measures and shot-noise rescaling values are available at \supplementarylink{} \cite{supplementary_material}.

\bibliographystyle{JHEP}
\bibliography{references,DESI2024,software}



\appendix

\section{Estimators for multi-tracer covariance matrices}
\label{sec:cov-estimation-extra}

In this section, we provide the generalized multi-tracer expressions for reference.
We have not endeavored to validate them in this work because in the \ezmocks{} suite different tracers have mostly been made from boxes at different redshifts.
We expect that the shift of the galaxies' positions with time will not allow self-consistent cross-correlations in such circumstances.
DESI has primarily focused on the LRG and ELG overlap in $z=0.8-1.1$ bin \cite{KP4s5-Valcin}, but \ezmocks{} for these tracers have been made from box snapshots at redshifts 1.1 and 0.95 respectively.

We add one capital Latin superscript to all the overdensities to signify the tracer involved (e.g. $\delta^X_i$).
For counts, it is two letters accordingly (e.g. $N^XN^Y$).

\subsection{Angular bins}
\label{sec:cov-estimation-multi-smu}

The Landy-Szalay estimator for the multi-tracer correlation function is
\begin{equation} \label{eq:Landy-Szalay-binned-multi}
\qty(\hat\xi^{XY})_a^c = \frac{\qty(N^XN^Y)_a^c}{\qty(R^XR^Y)_a^c}.
\end{equation}
We make the following shorthand for the multi-tracer covariance:
\begin{equation} \label{eq:cov2x2_def-multi}
\qty(C^{XY,ZW})_{ab}^{cd} \equiv \cov \qty[\qty(\hat\xi^{XY})_a^c, \qty(\hat\xi^{ZW})_b^d].
\end{equation}

It is important to note that the shot-noise approximation works only for the same-tracer overdensities:
\begin{equation} \label{eq:shot-noise-approximation-multi}
\qty(\delta^X_i)^2 \approx \frac{\snrescaling^X}{n^X_i} \qty(1+\delta^X_i).
\end{equation}

The full multi-tracer expression for the model covariance is from \cite{rascalC} but with some changes in notation:
\begin{align} \label{eq:Cov2x2estimator-multi}
\qty(\tilde C^{XY,ZW})_{ab}^{cd} \qty(\snrescalingvec) = \qty(^4C^{XY,ZW})_{ab}^{cd} + \frac{\snrescaling^X}4 \qty[\delta^{XW} \qty(^3C^{X,YZ})_{ab}^{cd} + \delta^{XZ} \qty(^3C^{X,YW})_{ab}^{cd}] \\ \nonumber
+ \frac{\snrescaling^Y}4 \qty[\delta^{YW} \qty(^3C^{Y,XZ})_{ab}^{cd} + \delta^{YZ} \qty(^3C^{Y,XW})_{ab}^{cd}] \\ \nonumber
+ \frac{\snrescaling^X \snrescaling^Y}2 \qty[\delta^{XW} \delta^{YZ} + \delta^{XZ} \delta^{YW}] \qty(^2C^{XY})_{ab}^{cd}
\end{align}
with
\begin{align} \label{eq:Cov2x2_234_Point_Defs-multi}
\qty(^4C^{XY,ZW})_{ab}^{cd} &= \frac{1}{\qty(R^XR^Y)_a^c \qty(R^ZR^W)_b^d} \sum_{i\neq j\neq k\neq l} n^X_i n^Y_j n^Z_k n^W_l w^X_i w^Y_j w^Z_k w^W_l \Theta^a(r_{ij}) \\ \nonumber
& \times \Theta^c(\mu_{ij}) \Theta^b(r_{kl}) \Theta^d(\mu_{kl}) \qty[\cancel{\eta^{({\rm c}),XYWZ}_{ijkl}} + \xi^{XZ}_{ik} \xi^{YW}_{jl} + \xi^{XW}_{il} \xi^{YZ}_{jk}] \\ \nonumber
\qty(^3C^{Y,XZ})_{ab}^{cd} &= \frac{4}{\qty(R^XR^Y)_a^c \qty(R^YR^Z)_b^d} \sum_{i\neq j\neq k} n^X_i n^Y_j n^Z_k w^X_i \qty(w^Y_j)^2 w^Z_k \Theta^a(r_{ij}) \Theta^c(\mu_{ij}) \\ \nonumber
& \times \Theta^b(r_{jk}) \Theta^d(\mu_{jk}) \qty[\cancel{\zeta^{XYZ}_{ijk}} + \xi^{XZ}_{ik}] \\ \nonumber
\qty(^2C^{XY})_{ab}^{cd} &= \frac{2\delta^{ab}\delta^{cd}}{\qty(R^XR^Y)_a^c \qty(R^XR^Y)_b^d}\sum_{i\neq j} n^X_i n^Y_j \qty(w^X_i w^Y_j)^2 \Theta^a(r_{ij}) \Theta^c(\mu_{ij}) \qty[1+\xi^{XY}_{ij}],
\end{align}
where $\delta^{XY}$, $\delta^{ab}$ and $\delta^{cd}$ are Kronecker deltas; $\xi^{XY}_{ij} = \xi^{XY}\qty(r_{ij}, \mu_{ij})$ is the 2PCF of tracers $X$ and $Y$ evaluated at the separation between points number $i$ and $j$ (in practice the value is obtained by bicubic interpolation from the input grid of correlation function values).

Analogously, $\zeta^{XYZ}_{ijk}$ and $\eta^{{(\rm c)},XYZW}_{ijkl}$ are the 3-point and connected 4-point correlation functions of the tracers listed in the superscript evaluated at the separations between $i,j,k$ and $i,j,k,l$ points, respectively.
These non-Gaussian higher-point functions are included for completeness but dropped in practice, we reflected this by crossing them out in the expressions.

The expressions for multi-tracer jackknife covariance can be constructed similarly.
However, to the best of our knowledge, only the single-tracer jackknife auto-covariances have been used in practice.
The shot-noise rescaling values have been tuned on each tracer separately.
Therefore the computation of the jackknife model for the other blocks of the full matrix has been unnecessary; the single-tracer expressions are sufficient.

\subsection{Projected Legendre}
\label{sec:cov-estimation-multi-legendre-projected}

Note that the Legendre projection factors (\cref{eq:Legendre-projection-factors}) are the same for all tracers.
Accordingly, the multi-tracer covariance is still projected linearly from the angularly binned one:
\begin{equation} \label{eq:full-cov-projection-Legendre-multi}
    \qty(\tilde C^{XY,ZW})_{ab}^{\ell\ell'} \equiv \cov \qty[\qty(\hat \xi^{XY})^\ell_a, \qty(\hat \xi^{ZW})^{\ell'}_b] = \sum_{c,d} \qty(\tilde C^{XY,ZW})_{ab}^{cd} F^\ell_c F^{\ell'}_d.
\end{equation}
Therefore the full covariance model is constructed analogously to \cref{eq:Cov2x2estimator-multi}:
\begin{align} \label{eq:cov2x2_Legendre-multi}
\qty(C^{XY,ZW})_{ab}^{\ell\ell'} = \qty(^4C^{XY,ZW})_{ab}^{\ell\ell'} + \frac{\snrescaling^X}{4} \qty[\delta^{XW} \qty(^3C^{X,YZ})_{ab}^{\ell\ell'} + \delta^{XZ} \qty(^3C^{X,YW})_{ab}^{\ell\ell'}] \\ \nonumber
+ \frac{\snrescaling^Y}{4} \qty[\delta^{YW} \qty(^3C^{Y,XZ})_{ab}^{\ell\ell'} + \delta^{YZ} \qty(^3C^{Y,XW})_{ab}^{\ell\ell'}] \\ \nonumber
+ \frac{\snrescaling^X \snrescaling^Y}{2} \qty(\delta^{XW} \delta^{YZ} + \delta^{XZ} \delta^{YW}) \qty(^2C^{XY})_{ab}^{\ell\ell'}.
\end{align}
with the following terms (similar to \cref{eq:Cov2x2_234_Point_Defs_Legendre_Projected}):
\begin{align} \label{eq:Cov2x2_234_Point_Defs_Legendre_Projected-multi}
\qty(^4C^{XY,ZW})_{ab}^{\ell\ell'} &= \sum_{i\neq j\neq k\neq l} n^X_i n^Y_j n^Z_k n^W_l w^X_i w^Y_j w^Z_k w^W_l \Theta^a(r_{ij}) \Theta^b(r_{kl}) \\ \nonumber
& \times \qty[\cancel{\eta^{({\rm c}),XYWZ}_{ijkl}} + 2\xi^{XZ}_{ik} \xi^{YW}_{jl}] \sum_c \frac{\Theta^c(\mu_{ij}) F^\ell_c}{\qty(R^XR^Y)_a^c} \sum_d \frac{\Theta^d(\mu_{kl}) F^{\ell'}_d}{\qty(R^ZR^W)_b^d}, \\ \nonumber
\qty(^3C^{Y,XZ})_{ab}^{\ell\ell'} &= 4 \sum_{i\neq j\neq k} n^X_i n^Y_j n^Z_k w^X_i \qty(w^Y_j)^2 w^Z_k \Theta^a(r_{ij}) \Theta^b(r_{jk}) \qty[\cancel{\zeta^{XYZ}_{ijk}} + \xi^{XZ}_{ik}] \\ \nonumber
& \times \sum_c \frac{\Theta^c(\mu_{ij}) F^\ell_c}{\qty(R^XR^Y)_a^c} \sum_d \frac{\Theta^d(\mu_{jk}) F^{\ell'}_d}{\qty(R^YR^Z)_b^d}, \\ \nonumber
\qty(^2C^{XY})_{ab}^{\ell\ell'} &= 2\delta^{ab} \sum_{i\neq j} n^X_i n^Y_j \qty(w^X_i w^Y_j)^2 \Theta^a(r_{ij}) \qty[1+\xi^{XY}_{ij}] \sum_c \frac{\Theta^c(\mu_{ij}) F^\ell_c F^{\ell'}_c}{\qty[\qty(R^XR^Y)_a^c]^2}.
\end{align}

The multi-tracer covariance model for Legendre multipoles of the 2PCF is implemented in the code and has been employed for overlapping tracers' cross-correlation in \cite{KP4s5-Valcin}.
We omit the corresponding jackknife expressions as they have not been used.

\section{Covariance for the combination of two regions}
\label{sec:cov-comb}

In this section, we document the procedure to compute the covariance matrix for the combination of two (or more) disjoint regions (volumes) using the covariances in each region.
It is fully consistent with the correlation function treatment in the DESI data processing pipeline.

\subsection{Angular bins}

During the combination of regions labeled by $G$, DESI scripts simply add the total counts of every relevant kind $QQ$ (where each $Q$ can be $D$ or $R$, data or randoms\footnote{I.e. $Q^XQ^Y$ encompasses $D^XD^Y$, $D^XR^Y$, $R^XD^Y$ and $R^XR^Y$ count arrays if standard BAO reconstruction is not used. After reconstruction we also have $S$ -- shifted randoms, and the counts relevant to the Landy-Szalay estimator are $D^XD^Y$, $D^XS^Y$, $S^XD^Y$, $S^XS^Y$ and $R^XR^Y$.}):
\begin{equation}
    \qty(Q^XQ^Y)_a^c = \sum_G \qty(Q^XQ^Y_G)_a^c.
\end{equation}

The following technical details are important for the strictly correct implementation with \pycorr{}.
The counts entering the correlation function estimator (\cref{eq:Landy-Szalay-binned-multi}) need to be normalized properly.
Originally the main cause was the possibly different number of galaxies and random points \cite{Landy-Szalay,2PCF-estimators-comparison}.
Now we are often weighting both types of points, and the pair counts weighted by the product of the two individual point weights can be normalized by the product of the sums of the weights\footnote{Exactly $\sum_i w_i^X \times \sum_j w_j^Y$ for cross-correlations ($X\ne Y$) and $\qty(\sum_i w_i^X)^2 - \sum_i \qty( w_i^X)^2$ for auto-correlations.}.
\pycorr{} stores both the weighted counts ({\tt wcounts}) and their normalization ({\tt wnorm}).
It is their ratio, the normalized counts, that enter the correlation function estimator (\cref{eq:Landy-Szalay-binned-multi}).
Before adding the pair counts for the two regions, DESI scripts bring counts of different types to the same normalization within each region (via the {\tt normalize} method\footnote{By default, this method preserves the {\tt wnorm} of $D^XD^Y$ counts.}).
During the combination of regions {\tt wnorm} are added as well as {\tt wcounts}.
Therefore the change of normalization within each region alters the relative contributions of the regions to different types of counts, even though it does not affect the region's correlation function.
Therefore this step is important to reproduce.
With all count types brought to the same normalization (in each region and therefore in their combination), the {\tt wnorm} cancel out in the Landy-Szalay estimator (\cref{eq:Landy-Szalay-binned-multi}) and so it is sufficient to substitute only {\tt wcounts}.

Then the total correlation function (\cref{eq:Landy-Szalay-binned-multi}) is simply the average of the two regions' correlation functions weighted by $RR$ counts:
\begin{align}
    \qty(\xi^{XY})_a^c &= \frac1{\qty(R^XR^Y)_a^c} \sum_G \qty(R^XR^Y_G)_a^c \qty(\xi^{XY}_G)_a^c \equiv \sum_G \qty(W^{XY}_G)_a^c \qty(\xi^{XY}_G)_a^c, \\
    \qty(W^{XY}_G)_a^c &\equiv \frac{\qty(R^XR^Y_G)_a^c}{\qty(R^XR^Y)_a^c}.
\end{align}
Then the covariance matrix for the combined region is simply
\begin{align}
    \qty(C^{XY,ZW})_{ab}^{cd} \approx \sum_G \qty(C^{XY,ZW}_G)_{ab}^{cd} \qty(W^{XY}_G)_a^c \qty(W^{ZW}_G)_b^d,
\end{align}
where we neglect the covariance between the correlation functions in different regions, the estimation of which poses extra challenges.
We expect this to be a safe approximation for the North and South Galactic caps in the DESI footprint because the separation between galaxies in different caps is significantly larger than the maximum separation for correlation function measurements (200 $\ihMpc$).

\subsection{Legendre}

We can build upon the results for angular bins, with conversions both ways between them and Legendre moments.
The Legendre multipoles are estimated from the angular bins via \cref{eq:Legendre-from-binned-2PCF,eq:Legendre-projection-factors}:
\begin{equation}
    \qty(\xi^{XY})^\ell_a = \sum_c \qty(\xi^{XY})_a^c F^\ell_c.
\end{equation}
One can do the reverse approximately with bin-averaged values of the Legendre polynomials\footnote{Or possibly with just the middles of the bins, but this way they end up very much related to already computed projection factors $F^\ell_c$.}:
\begin{align}
    L_\ell^c &\equiv \frac1{\Delta\mu_c} \int_{\Delta\mu_c} d\mu\, L_\ell(\mu) = \frac{F^\ell_c}{(2\ell + 1) \Delta\mu_c} \label{eq:Legendre-polynomial-means}, \\
    \qty(\xi^{XY})_a^c &\approx \sum_\ell \qty(\hat \xi^{XY})^\ell_a L_\ell^c. \label{eq:binnned-from-Legendre-2PCF}
\end{align}
These do not depend on the region.
Additional approximation comes from the fact that we limit the multipole index --- in this work, we only consider $\ell=0,2,4$.

With this we can work out the partial derivative of the Legendre moment of the combined correlation function with respect to one in each of the regions in the following steps:
\begin{equation}
    \frac{\partial \qty(\xi^{XY})^\ell_a}{\partial \qty(\xi^{XY}_G)^{\ell_1}_a} = \sum_c \frac{\partial \qty(\xi^{XY})^\ell_a}{\partial \qty(\xi^{XY})^c_a} \frac{\partial \qty(\xi^{XY})^c_a}{\partial \qty(\xi^{XY}_G)^c_a} \frac{\partial \qty(\xi^{XY}_G)^c_a}{\partial \qty(\xi^{XY}_G)^{\ell_1}_a} \approx \sum_c F_c^\ell \qty(W^{XY}_G)_a^c L^c_{\ell_1} \equiv \qty(W_G^{XY})^{\ell,\ell_1}_a;
\end{equation}
the different separation bins and correlation functions stay independent.

Then the covariance matrix for the combined region's 2PCF is
\begin{align}
    \qty(C^{XY,ZW})_{ab}^{\ell\ell'} \approx \sum_G \sum_{\ell_1,\ell'_1} \qty(C^{XY,ZW}_G)_{ab}^{\ell_1\ell'_1} \qty(W^{XY}_G)_a^{\ell,\ell_1} \qty(W^{ZW}_G)_b^{\ell',\ell'_1},
\end{align}
here we also neglect the covariance between the correlation functions in different regions.

\section{Author affiliations}
\label{sec:affiliations}

\noindent \hangindent=.5cm $^{1}${Center for Astrophysics $|$ Harvard \& Smithsonian, 60 Garden Street, Cambridge, MA 02138, USA}

\noindent \hangindent=.5cm $^{2}${Institute of Physics, Laboratory of Astrophysics, \'{E}cole Polytechnique F\'{e}d\'{e}rale de Lausanne (EPFL), Observatoire de Sauverny, Chemin Pegasi 51, CH-1290 Versoix, Switzerland}

\noindent \hangindent=.5cm $^{3}${IRFU, CEA, Universit\'{e} Paris-Saclay, Road D36, 91191 Gif sur Yvette, France}

\noindent \hangindent=.5cm $^{4}${Physics Department, Yale University, P.O. Box 208120, New Haven, CT 06511, USA}

\noindent \hangindent=.5cm $^{5}${Department of Physics \& Astronomy, Ohio University, 139 University Terrace, Athens, OH 45701, USA}

\noindent \hangindent=.5cm $^{6}${Center for Cosmology and AstroParticle Physics, The Ohio State University, 191 West Woodruff Avenue, Columbus, OH 43210, USA}

\noindent \hangindent=.5cm $^{7}${Department of Astronomy, The Ohio State University, 4055 McPherson Laboratory, 140 W 18th Avenue, Columbus, OH 43210, USA}

\noindent \hangindent=.5cm $^{8}${Lawrence Berkeley National Laboratory, 1 Cyclotron Road, Berkeley, CA 94720, USA}

\noindent \hangindent=.5cm $^{9}${Physics Dept., Boston University, 590 Commonwealth Avenue, Boston, MA 02215, USA}

\noindent \hangindent=.5cm $^{10}${University of Michigan, 500 S. State Street, Ann Arbor, MI 48109, USA}

\noindent \hangindent=.5cm $^{11}${Leinweber Center for Theoretical Physics, University of Michigan, 450 Church Street, Ann Arbor, Michigan 48109-1040, USA}

\noindent \hangindent=.5cm $^{12}${Department of Physics \& Astronomy, University College London, Gower Street, London, WC1E 6BT, UK}

\noindent \hangindent=.5cm $^{13}${Institute for Computational Cosmology, Department of Physics, Durham University, South Road, Durham DH1 3LE, UK}

\noindent \hangindent=.5cm $^{14}${Instituto de F\'{\i}sica, Universidad Nacional Aut\'{o}noma de M\'{e}xico, Circuito de la Investigaci\'{o}n Cient\'{\i}fica, Ciudad Universitaria, Cd. de M\'{e}xico C.~P.~04510, M\'{e}xico}

\noindent \hangindent=.5cm $^{15}${Department of Astronomy, School of Physics and Astronomy, Shanghai Jiao Tong University, No.800 Dong Chuan Road, No.5 Physics building, Minhang District, Shanghai 200240, China}

\noindent \hangindent=.5cm $^{16}${Kavli Institute for Particle Astrophysics and Cosmology, Stanford University, 2575 Sand Hill Road, M/S 29, Menlo Park, CA 94025, USA}

\noindent \hangindent=.5cm $^{17}${SLAC National Accelerator Laboratory, 2575 Sand Hill Road, Menlo Park, CA 94025, USA}

\noindent \hangindent=.5cm $^{18}${University of California, Berkeley, 110 Sproul Hall \#5800 Berkeley, CA 94720, USA}

\noindent \hangindent=.5cm $^{19}${Institut de F\'{i}sica d’Altes Energies (IFAE), The Barcelona Institute of Science and Technology, Edifici Cn, Campus UAB, 08193 Bellaterra (Barcelona), Spain}

\noindent \hangindent=.5cm $^{20}${Departamento de F\'isica, Universidad de los Andes, Cra. 1 No. 18A-10, Edificio Ip, CP 111711, Bogot\'a, Colombia}

\noindent \hangindent=.5cm $^{21}${Observatorio Astron\'omico, Universidad de los Andes, Cra. 1 No. 18A-10, Edificio H, CP 111711 Bogot\'a, Colombia}

\noindent \hangindent=.5cm $^{22}${Department of Physics, The University of Texas at Dallas, 800 W. Campbell Rd., Richardson, TX 75080, USA}

\noindent \hangindent=.5cm $^{23}${Departament de F\'{\i}sica Qu\`{a}ntica i Astrof\'{\i}sica, Universitat de Barcelona, Mart\'{\i} i Franqu\`{e}s 1, E08028 Barcelona, Spain}

\noindent \hangindent=.5cm $^{24}${Institut d'Estudis Espacials de Catalunya (IEEC), Campus Diagonal Nord, Edifici Nexus II, C/ Jordi Girona, 29, 08034 Barcelona Barcelona, Spain}

\noindent \hangindent=.5cm $^{25}${Institut de Ci\`encies del Cosmos (ICCUB), Universitat de Barcelona (UB), c. Mart\'i i Franqu\`es, 1, 08028 Barcelona, Spain.}

\noindent \hangindent=.5cm $^{26}${Consejo Nacional de Ciencia y Tecnolog\'{\i}a, Av. Insurgentes Sur 1582. Colonia Cr\'{e}dito Constructor, Del. Benito Ju\'{a}rez C.P. 03940, M\'{e}xico D.F. M\'{e}xico}

\noindent \hangindent=.5cm $^{27}${Departamento de F\'{\i}sica, DCI-Campus Le\'{o}n, Universidad de Guanajuato, Loma del Bosque 103, Le\'{o}n, Guanajuato C.~P.~37150, M\'{e}xico}

\noindent \hangindent=.5cm $^{28}${Fermi National Accelerator Laboratory, PO Box 500, Batavia, IL 60510, USA}

\noindent \hangindent=.5cm $^{29}${Department of Physics, The Ohio State University, 191 West Woodruff Avenue, Columbus, OH 43210, USA}

\noindent \hangindent=.5cm $^{30}${School of Mathematics and Physics, University of Queensland, Physics Annexe (6), Level 2, St Lucia, Brisbane, QLD 4072, Australia}

\noindent \hangindent=.5cm $^{31}${NSF NOIRLab, 950 N. Cherry Ave., Tucson, AZ 85719, USA}

\noindent \hangindent=.5cm $^{32}${Sorbonne Universit\'{e}, CNRS/IN2P3, Laboratoire de Physique Nucl\'{e}aire et de Hautes Energies (LPNHE), 4 Pl. Jussieu, 75005 Paris, France}

\noindent \hangindent=.5cm $^{33}${Departament de F\'{i}sica, Serra H\'{u}nter, Universitat Aut\`{o}noma de Barcelona, Carretera de Bellaterra a l'Autònoma, 08193 Cerdanyola del Vallès, Barcelona, Spain}

\noindent \hangindent=.5cm $^{34}${Laboratoire de Physique Subatomique et de Cosmologie, 53 Avenue des Martyrs, 38000 Grenoble, France}

\noindent \hangindent=.5cm $^{35}${Instituci\'{o} Catalana de Recerca i Estudis Avan\c{c}ats, Passeig de Llu\'{\i}s Companys, 23, 08010 Barcelona, Spain}

\noindent \hangindent=.5cm $^{36}${Department of Physics and Astronomy, University of Sussex, Pevensey 2 Building, Falmer Campus, Brighton BN1 9QH, U.K}

\noindent \hangindent=.5cm $^{37}${Department of Physics \& Astronomy, University  of Wyoming, 1000 E. University, Dept.~3905, Laramie, WY 82071, USA}

\noindent \hangindent=.5cm $^{38}${National Astronomical Observatories, Chinese Academy of Sciences, A20 Datun Rd., Chaoyang District, Beijing, 100012, P.R. China}

\noindent \hangindent=.5cm $^{39}${Instituto Avanzado de Cosmolog\'{\i}a A.~C., San Marcos 11 - Atenas 202. Magdalena Contreras, 10720. Ciudad de M\'{e}xico, M\'{e}xico}

\noindent \hangindent=.5cm $^{40}${Department of Physics and Astronomy, University of Waterloo, 200 University Ave W, Waterloo, ON N2L 3G1, Canada}

\noindent \hangindent=.5cm $^{41}${Waterloo Centre for Astrophysics, University of Waterloo, 200 University Ave W, Waterloo, ON N2L 3G1, Canada}

\noindent \hangindent=.5cm $^{42}${Perimeter Institute for Theoretical Physics, 31 Caroline St. North, Waterloo, ON N2L 2Y5, Canada}

\noindent \hangindent=.5cm $^{43}${Space Sciences Laboratory, University of California, Berkeley, 7 Gauss Way, Berkeley, CA  94720, USA}

\noindent \hangindent=.5cm $^{44}${Max Planck Institute for Extraterrestrial Physics, Gie\ss enbachstra\ss e 1, 85748 Garching, Germany}

\noindent \hangindent=.5cm $^{45}${Department of Physics, Kansas State University, 116 Cardwell Hall, Manhattan, KS 66506, USA}

\noindent \hangindent=.5cm $^{46}${Department of Physics and Astronomy, Sejong University, 209 Neungdong-ro, Gwangjin District, Seoul, 143-747, Korea}

\noindent \hangindent=.5cm $^{47}${Centre for Astrophysics \& Supercomputing, Swinburne University of Technology, P.O. Box 218, Hawthorn, VIC 3122, Australia}

\noindent \hangindent=.5cm $^{48}${CIEMAT, Avenida Complutense 40, E-28040 Madrid, Spain}

\noindent \hangindent=.5cm $^{49}${Department of Physics, University of Michigan, 450 Church Street, Ann Arbor, MI 48109, USA}

\noindent \hangindent=.5cm $^{50}${Department of Astronomy, Tsinghua University, 30 Shuangqing Road, Haidian District, Beijing, China, 100190}


\end{document}